\documentclass[12pt,authoryear]{article}
\usepackage[utf8]{inputenc} 
\usepackage[T1]{fontenc}    
\usepackage{url}            
\usepackage{booktabs}       
\usepackage{amsmath}
\usepackage{amsfonts}       
\usepackage{nicefrac}       
\usepackage{lmodern}  

\usepackage{microtype}      
\usepackage{graphicx}
\usepackage{graphics}
\usepackage{bm}
\usepackage[round]{natbib}
\usepackage{float}
\usepackage{color, colortbl}
\usepackage{algorithm}
\usepackage{algpseudocode}
\usepackage{enumerate}
\newcommand{\blind}{1}

\begin{document}

\newcommand{\bx}{\boldsymbol{x}}

\def\spacingset#1{\renewcommand{\baselinestretch}%
{#1}\small\normalsize} \spacingset{1}


\if1\blind
{
  \title{\bf Loss-based prior for CART and BART models}
  \author{Francesco Serafini\hspace{.2cm}\\
    School of Earth Sciences, University of Bristol\\
    and \\
    Cristiano Villa \\
    School of Mathematics, Duke Kunshan University \\
    and \\
    Fabrizio Leisen \\
    School of Mathematics, Kings College London \\
    and \\
    Kevin Wilson \\
    School of Mathematics, Statistics and Physics, Newcastle Univeristy }
  \maketitle
} \fi

\if0\blind
{
  \bigskip
  \bigskip
  \bigskip
  \begin{center}
    {\LARGE\bf Loss-based prior for  BART models}
\end{center}
  \medskip
} \fi

\bigskip
\begin{abstract}
We present a novel prior for tree topology within Bayesian Additive Regression Trees (BART) models. This approach quantifies the hypothetical loss in information and the loss due to complexity associated with choosing the “wrong” tree structure. The resulting prior distribution is compellingly geared toward sparsity — a critical feature considering BART models’ tendency to overfit. Our method incorporates prior knowledge into the distribution via two parameters that govern the tree's depth and balance between its left and right branches. Additionally, we propose a default calibration for these parameters, offering an objective version of the prior. We demonstrate our method’s efficacy on both simulated and real datasets.
\end{abstract}

\noindent%
\vfill
\newpage

\spacingset{1.45} 
\section{Introduction}
\label{sec:intro}
\emph{Bayesian Classification and Regression Tree} \citep[CART, ][]{chipman1998bayesian, denison1998bayesian} and its generalisation,  
\emph{Bayesian additive regression trees} \citep[BART, ][]{chipman2006bayesian, chipman2010bart}, form a flexible class of semi-parametric models for regression and classification problems, especially useful in the presence of a high number of covariates. The main objective of BART is to model an unknown function, linking the covariates to the observations, as a sum of binary $m$ regression trees \citep{breiman2017classification}, CART does the same but considering $m = 1$. BART models were firstly introduced for regression problems with Gaussian errors and multicategory classification problems but, since then, they have been generalised to multinomial and multinomial logit models \citep{murray2017log}, Gaussian models with variance depending on covariates \citep{pratola2020heteroscedastic}, count data \citep{murray2017log}, gamma regression models \citep{linero2020semiparametric}, and to Poisson processes \citep{lamprinakou2023bart}. They also have been adapted to face different problems, such as variable selection \citep{linero2018bayesian}, regression with monotonicity constraints \citep{chipman2022mbart}, survival analysis \citep{bonato2011bayesian, sparapani2016nonparametric}, and causal inference \citep{hill2011bayesian, hahn2020bayesian} among others. Furthermore, much effort has been devoted in building a theoretical framework for BART to study posterior convergence \citep{rockova2017posterior, rovckova2019theory, rockova2019semi, jeong2023art}. We refer to \cite{hill2020bayesian} for a thorough introduction to BART models and their applications.

One characteristic of BART and CART models is that the prior on the tree space acts as a regularisation prior. Specifically, the prior is defined by specifying a distribution on the binary tree space and on the splitting rules at the internal nodes, and a conditional distribution on the value at the terminal nodes. The prior on the tree space is used to downweight \emph{undesirable} trees according to specific characteristics measuring tree complexity (e.g. number of terminal nodes, depth). This is because it is desirable to avoid situations in which one tree is unduly influential, and to advantage cases where each tree captures a specific aspect of the data. The most used tree prior is the one proposed by \cite{chipman1998bayesian} which specifies, for each node, the probability that the node is a split as a decreasing function of the node's depth, and the tree prior is the product of the nodes' probabilities. We refer to this prior as the \emph{classic tree prior} (CL) as it is the most used in practice. While intuitively appealing, using this prior makes relatively difficult to access the distribution of quantities such as the number of terminal nodes, or the depth, and to set parameters of the prior to have the desired prior probabilities. \cite{denison1998bayesian} proposed an alternative by directly specifying a distribution on the number of terminal nodes, and a uniform distribution over the trees with a given number of terminal nodes. However, this prior tends to concentrate around skewed trees with a large difference between terminal nodes on the left and right branch. \cite{wu2007bayesian} introduced the \emph{pinball prior} in which they mix these approaches by considering a prior on the number of terminal nodes and, cascading down from the root of the tree, a probability on the number of terminal nodes going left and right. This gives control over the \emph{shape} of tree being more or less skewed, while the CL prior only favours balanced trees with the same number of terminal nodes on the left and right branch. However, as for the CL prior, the distribution of the number of terminal nodes, the depth, or the skewness, is accessible only through simulations, and it is difficult to quantify how much prior probabilities will be affected by varying the prior parameters. Other examples that aim to overcome the problem are the \emph{spike-and-tree} \citep{rockova2017posterior} and the Dirichlet prior \citep{linero2018bayesian} which both penalise the complexity of the tree by specifying a sparse prior on the predictors' space. For all the priors mentioned above the choice of parameters is subjective and researchers usually rely on default values lacking mathematical (and objective) motivations.

A way to design an objective prior distribution is the loss-based prior approach developed by \cite{villa2015objective}. This is based on considering the prior for a tree to be proportional to a function of the loss incurred when selecting a different tree than the one generating the data. The loss is considered both in terms of information and in terms of complexity. The prior obtained with this approach is appealing for BART and CART models because it explicitly penalises for the complexity of the tree. Furthermore, the parameters of the prior distribution can be chosen by maximising the expected loss and therefore it is mathematically justified and objective. The loss-based approach has been used to design objective priors in a variety of contexts: for the parameters of a standard, skewed and multivariate $t$-distribution \citep{villa2014, leisen2017objective, villa2018objective}, for discrete parameter spaces \citep{villa2015objectivedis}, for time series analysis \citep{leisen2020loss}, for change-point analysis \citep{hinoveanu2019bayesian}, for the number of components in a mixture model \citep{grazian2020loss}, for variable selection in linear regression \citep{villa2020variable}, and for Gaussian graphical models \citep{hinoveanu2020loss}. 

In this article, we apply the loss-based approach to design an objective prior distribution for the tree structure in BART and CART models. The loss-based (LB) prior we propose depends on the number of terminal nodes and the difference between the number of terminal nodes on the left and right branch (used as a measure of skewness) offering control over these two quantities. We remark that this in not the only possible choice, and the approach we are presenting can be tailored to different needs. We chose this two tree statistics because also the CL and the pinball prior indirectly penalise for these. One of the advantages is that the distribution of the number of terminal nodes and the left and right difference are directly available, the relationship with prior parameters is intuitive, and therefore it is easier to calibrate them. In absence of prior information, the LB approach allows to calibrate the parameters of the prior by maximising the expected loss providing an objective way to determine a default distribution. Furthermore, from our synthetic (on CART) and real data (two: one on CART, one on BART) experiments, it turns out that, considering the same expected number of terminal nodes \emph{a priori}, the loss-based prior explores shallower, less complex trees, than the CL prior without losing in terms of performance. This is relevant, especially for variable selection, because it means that the loss-based prior provides a greater penalty to increase complexity which stimulates more competition among predictors to enter the pool, and therefore provides more meaningful predictors set. In all cases, we found that the loss-based  prior with parameters calibrated via expected loss maximisation is the one providing the best trade-off between tree complexity and goodness-of-fit.

The prior we propose in this article has other useful features. Computationally, the LB prior needs only the number of terminal nodes ($n_L$) and the difference between left ($n_l$) and right ($n_r$) terminal nodes ($\Delta = |n_l - n_r|$). As is done for the CL prior, these quantities can be tracked and updated during MCMC iterations, so to reduce the computational cost of calculating the prior. Theoretically, using the loss-based prior approach, we find a prior distribution for a tree $T$ which is simple, and has an intuitive meaning,

\begin{equation}
\pi(T) \propto \exp\{-\omega n_L(T) - \gamma \Delta(T)\}
\end{equation}
where $\omega, \gamma$ are parameters. We explain how this is derived in Section \ref{sec:lbprior_BART}. The loss-based prior approach is flexible and can be used to design different priors to the one we present here. Indeed, one can use the same methodology applied here but considering different measures of tree statistics tailored to the problem. In the same spirit, one can use the LB approach to design priors for other quantities. For example, the number of trees $m$ in BART models is a quantity for which there is currently no default prior, and it is chosen by trying out different values. In future works, the LB approach could be extended to this quantity, thus providing a unified approach to set the prior on the tree complexity and the number of trees in BART models.

The paper is organised as follows: Section \ref{sec:prel} introduces formally the BART model, the prior proposed by \citep{chipman1998bayesian} for the tree structure and the loss-based approach. Section \ref{sec:lbprior_BART} describes the prior on the tree structure obtained using the loss-based approach. Section \ref{sec:MCMC} describes the MCMC algorithm used to explore the posterior distribution. Section \ref{sec:sim_study} compares the performance of the CL prior with the LB prior obtained in Section \ref{sec:lbprior_BART} on simulated data. The comparison includes instances of the LB prior replicating the default CL prior, and vice versa. Section \ref{sec:real_data} compares the performance of the default classic prior and the default LB prior on the breast cancer data analysed in \cite{chipman1998bayesian} and \cite{wu2007bayesian} and the diabetes data provided by \cite{diabetes}. This is the first time a BART model is used to study this diabetes dataset. Finally, in Section \ref{sec:conclusions} we discuss the results shown in the article and draw conclusions.

\section{Preliminaries on BART and loss-based approach}
\label{sec:prel}

In this Section, we set up the notation used through the paper, give a formal description of the BART model, the CL prior proposed by \cite{chipman1998bayesian}, and the loss-based prior approach that is used in Section \ref{sec:lbprior_BART} to design an objective prior on the tree structure.

\subsection{BART models}

The BART model \cite[BART,][]{chipman2010bart} has the objective of making inference about an unknown function $f(\bx)$ that predicts an output $Y$ such that, in the case of gaussian residuals,

\begin{equation*}
Y = f(\bx) + \epsilon, \quad\quad \epsilon \sim N(0, \sigma^2), 
\end{equation*}
where $\bx = (x_1,\dots,x_p)$ is a $p$-dimensional vector of covariates, usually assumed to be $\bx \in [0,1]^p$. The function $f(\cdot)$ is unknown and assumed to be smooth. The idea is to approximate $f(\cdot)$ using a sum of $m \geq 1$ regression trees so that 

\begin{equation}
Y = \sum_{j = 1}^m g(\bx, T_j, M_j) + \epsilon, \quad\quad \epsilon \sim N(0, \sigma^2),
\label{eq:BART_def}
\end{equation}
where $T_j$ is a binary tree composed by splitting rules of the kind $x_i \leq \tau$ at each internal node, $M_j = (\mu_1,\dots,\mu_{n_L(T_j)})$ contains the values at the terminal nodes, $n_L(T_j)$ is the number of terminal nodes, and $g(\bx, T_j, M_j)$ represents the $j$-th regression tree. CART models are the same but strictly considering $m = 1$.

Essentially, a regression tree is a way to represent a piecewise constant function on a partition of the domain. The internal nodes of the tree represent the partition, the terminal nodes represent the different elements of the partition, and the value at the terminal nodes represents the value assumed by the function on the corresponding element of the partition. However, regression trees are not capable of representing any partition, but only partitions composed by non-overlapping rectangles. In other words, we are interested in partitions obtained by nested parallel-axis splits. 

Each internal node is equipped with a splitting rule on one of the predictor directions of the form $x_i \leq \tau$ for $i = 1,...,p$. So, a splitting rule is composed by a splitting variable $i$ and a splitting value $\tau$. If the observation meets the splitting rule, we move on the left branch of the tree, otherwise we move on the right branch. In this way, each observation ($\bx$) is associated with a terminal node of the tree ($f(\bx)$). Therefore, the values at the terminal nodes depend on which observations are associated with each terminal node. 

In order to be able to estimate the values at the terminal nodes, we are mostly interested in partitions with at least one (or $n_\mu$) observation associated with each terminal node. \cite{chipman1998bayesian} call these partitions \emph{valid}, and this property depends on the available data. For example, if we have $n$ observations, and we want at least one observation per terminal node, then, the maximum number of terminal node of a valid tree is $n$. In this article, we only focus on \emph{valid} trees, a formal definition is given in Appendix A.

\subsection{Priors for BART}
\label{sec:classic_prior}
In both CART and BART models, the same prior is usually considered. For a single tree regression model, we need to define a prior on the tree topology (the shape of the tree), the splitting rules, the values at the terminal nodes, and the additional parameters needed to calculate the likelihood (e.g. marginal variance for Gaussian regression). These priors are usually assumed to be independent \citep{chipman1998bayesian} so the prior for the whole set of parameters is simply the product of the above priors. Given that in this article we provide a new prior for the tree topology, in this section we only describe said prior. Furthermore, in BART models, the same prior is used for all trees, so that it is sufficient to specify it for one. We assume through out the paper that the prior on the splitting rules, the values at the terminal nodes, and the remaining likelihood parameters, are the same as described in \cite{chipman1998bayesian} and \cite{chipman2010bart}.

In CART/BART models, the prior on the tree topology plays the role of a \emph{regularisation} prior, meaning that it acts as a penalty on the tree complexity. This is needed in order to avoid overfitting and to keep the relative contribution of each tree to the summation balanced. Should the prior not penalise for complexity, there is a high probability of having one complex and rich tree, while all the others would have a negligible contribution to the summation. This limits interpretability and increases the chances of overfitting. Furthermore, having shallower trees increases the competition between predictors to enter the pool, which in turns encourages only the most important variables to be selected.

The most widely used regularisation prior for the tree topology is the one originally proposed by \cite{chipman1998bayesian} for a CART models. Through out this article, we refer to this prior as the \emph{classic tree prior} (CL). The CL prior is defined by providing for each internal node the probability that the node is a split. For a node at depth $d$ (the minimum number of steps from the root to the node) the probability that the node is a split is given by

\begin{equation*}
\alpha(1 + d)^{-\beta}, \quad \alpha \in (0,1), \beta \geq 0.
\end{equation*}

Given a binary tree $T$ with internal nodes index set $a(T)$ and terminal nodes index set $b(T)$ the CL prior for tree $T$ is then given by

\begin{equation*}
\pi_C(T) = \prod_{i \in a(T)} \alpha(1 + d_i)^{-\beta} \prod_{j \in b(T)} \left( 1 - \alpha(1 + d_j)^{-\beta} \right).
\end{equation*}  

The CL prior penalises for complexity assuming that the probability that a node is a split decreases with the depth of the tree. The prior is governed by two parameters $\alpha$ and $\beta$. Parameter $\alpha$ corresponds to the probability that the root node is a split (the first node has depth 0). Parameter $\beta$ regulates how fast the probability that a node is a split decays as a function of the node's depth. The default setting used in applications \citep{hill2011bayesian, zhang2020application, sparapani2020nonparametric} and in R-packages for BART models such as \texttt{dbarts} \citep{dbarts} are $\alpha = 0.95$ and $\beta = 2$. 

\begin{figure}
\centering
\includegraphics[width=\linewidth]{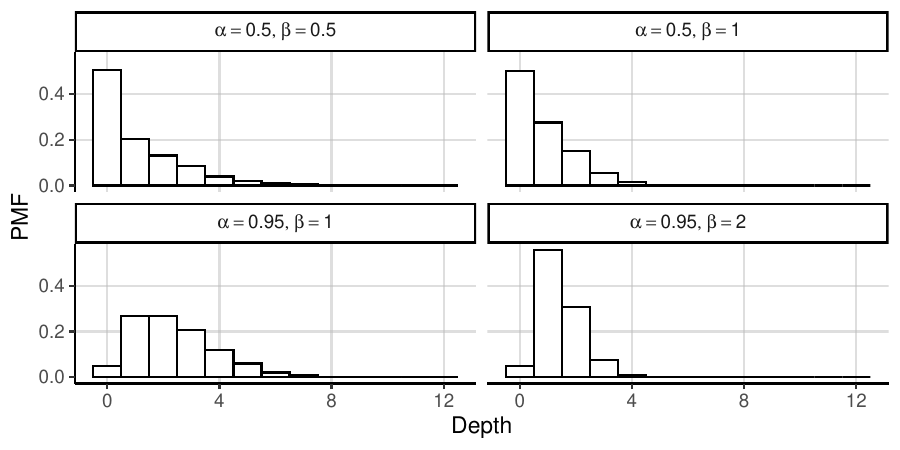}
\caption{Depth distribution under the CL prior for different combinations of prior parameters $\alpha$ and $\beta$ obtained simulating 10000 trees from the prior.}
\label{fig:1_depth_chipman}
\end{figure}

\begin{figure}
\centering
\includegraphics[width=\linewidth]{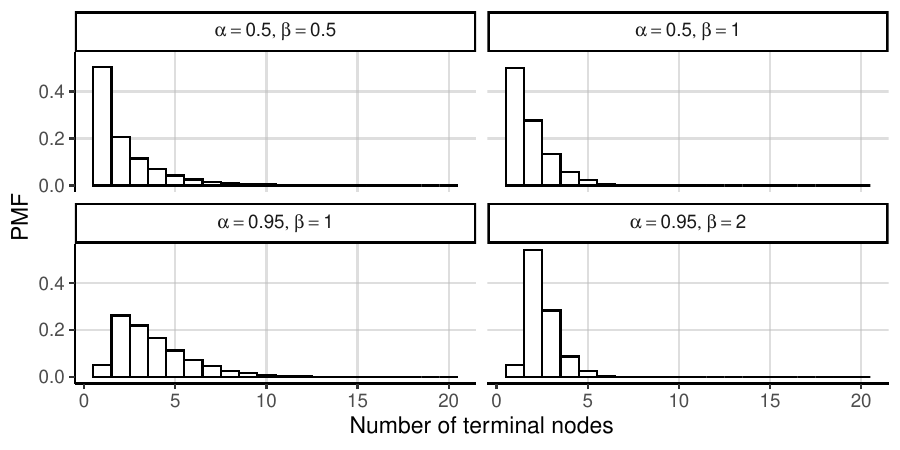}
\caption{Number of terminal nodes probability mass function under the CL prior for different combinations of prior parameters $\alpha$ and $\beta$ obtained simulating 10000 trees from the prior.}
\label{fig:2_nterm_chipman}
\end{figure}

The CL prior formulation induces a distribution on the space of the binary trees which penalises for complexity, in the sense that trees with more terminal nodes, or with terminal nodes at greater depths, are assigned less prior probability. This means that it also indirectly penalises skewed trees with more terminal nodes on one side. This is because, for two trees with the same number of terminal nodes, the one with the greatest difference between terminal nodes on the left and right branches is the one having splits at higher depth, and therefore it has a smaller prior probability. Under this prior, retrieving an analytical expression for the distribution on the number of terminal nodes or the depth of the tree (defined as the maximum depth of a node in the tree) is cumbersome, and the only way to retrieve these distributions is through simulation. Figures \ref{fig:1_depth_chipman} and \ref{fig:2_nterm_chipman} show the depth and number of terminal nodes distributions for different values of the prior parameters. 

The fact that the distributions of tree-related quantities under the CL prior are accessible only through simulation is a disadvantage of this prior. Indeed, this makes complicated calibrating the prior parameters by specifying the mean or quantiles of the number of terminal nodes or depth distributions because one have to find the right parameters value by trail and error. Also, it is unclear how varying the parameters affects other quantities indirectly penalises such as the difference between left and right terminal nodes. Researchers usually starts from the default and then tweak the parameters based on posterior results \citep{zhang2020application, hill2011bayesian}. This procedure is inefficient, introduce a degree of subjectivity in the analysis, and limits reproducibility. Instead, this problem vanishes when using our LB prior, both because the default parameter values are determined objectively, and because the distribution of the number of terminal nodes and the skewness of the tree are analytically available, and the effect on these of varying the prior parameters is clear and intuitive.

\subsection{Loss-based priors}
The loss-based approach \citep{villa2015objective} is a technique to design objective prior distributions that has been applied to different problems  \citep{villa2015objectivedis, leisen2020loss, grazian2020loss, hinoveanu2020loss}. The idea is that the choice of a “wrong” model yields to a loss, which has two components: one in information and one in complexity. The former stems from a well-known Bayesian property \citep{berk1966limiting}: if a model is misspecified, the posterior distribution will asymptotically accumulate on the model that is the most similar to the true one, where this similarity is measured in terms of the Kullback--Leibler divergence \citep[KLD,][]{kullback1951information}. The latter loss originates from the fact that the more complex a model is, the more components have to be considered and measured. Essentially, for a model $\mathcal M$ in a family $\mathcal F$, we have that 

\begin{equation*}
\pi(\mathcal M) \propto \exp\{Loss_I(\mathcal M) + Loss_C(\mathcal M)\},
\end{equation*}
where $Loss_I(\mathcal M)$ is the loss in information, which occurs by selecting a model different from $\mathcal M$ when $\mathcal M$ is the data-generating model, and $Loss_C(\mathcal M)$ is the loss in complexity incurred by selecting model $\mathcal M$. The key point is that models for which we incur a greater loss in information when misspecified should have more prior probability, and more complex models should be penalised in accordance with a parsimony principle. 

The loss in information for a model is considered to be the minimum KLD between the model and the alternative models in a pre-defined family $\mathcal F$. For a model $\mathcal M$, the greater the minimum KLD, the more difficult is to find an alternative model within $\mathcal F$ bringing similar information. We assign greater prior distribution to models with higher KLD because we would incur in a greater loss if misspecified.

More formally, given a variable of interest $Y \in \mathcal Y$, a set of predictors $\bx \in \mathcal X$, and two models $\mathcal M$ and $\mathcal M'$ in $\mathcal F$, with distributions $\pi(y|\bx)$ and $\pi'(y|\bx)$, the Kullback-Leibler divergence between the two models is 

\begin{equation*}
KL(\mathcal M || \mathcal M') = \int_{\mathcal Y} \pi(y| \bx) \log\left(\frac{\pi(y | \bx)}{\pi'(y | \bx)}\right)\,dy,
\label{eq:DKL_def}
\end{equation*}

and the loss of information is then given by

\begin{equation*}
Loss_I(M) = \min_{\mathcal M' \in \mathcal F} KL(\mathcal M||\mathcal M').
\end{equation*}

Contrarily, the loss in complexity strictly depends on the problem at hand and different choices are possible. For example, a natural choice in problems of variable selection is to consider the number of active predictors as loss in complexity \citep{villa2020variable}. This makes the loss-based prior approach adaptable to different problems.

\section{Loss-based prior for BART}
\label{sec:lbprior_BART}

In this Section, we show how to implement the loss-based approach to design a prior distribution for the tree topology of a BART model. As for the CL prior (see Section \ref{sec:classic_prior}) the prior is defined for a single tree and applied to all trees involved in the summation. In this context, model $\mathcal M$ is represented by the tree $T$, the distribution of the observations $\pi(y | \bx)$ is the distribution induced by the tree $\pi(y| \bx, T, M)$, and the family $\mathcal F$ is the set of binary trees. We will discuss separately the loss in information, the loss in complexity and the resulting prior.

\subsection{Loss in information}

The KLD between two trees $T$ and $T'$ is given by

\begin{equation*}
KL(T||T' | \bx, M, M') = \int_\mathcal Y \pi(y|\bx, T, M) \log\left(\frac{\pi(y|\bx, T, M)}{\pi(y|\bx, T', M')}\right)dy,
\end{equation*}
and the loss in information is given by
\begin{equation}
Loss_I(T) = \min_{T', M'} KL(T||T' | \bx, M, M').
\label{eq:LI}
\end{equation}

The KLD is zero when it is possible to find a tree $T'$ with terminal node values $M'$ that replicates exactly $T$ and $M$. In other words, if we can find $T', M'$ such that $g(\bx, T, M) = g(\bx, T', M')$ for each $\bx$, then the KLD is zero and there is no loss in information in misspecifying the model. This is similar to having nested models where the more complex model replicates the simpler ones. Without considering any limitation on the tree complexity (e.g. maximum number of terminal nodes, maximum depth, etc) it is always possible to find a tree $T' \neq T$ with terminal nodes values $M'$ that replicates $T, M$. Indeed, it is sufficient to consider $T'$ to be obtained by splitting one terminal node (say $j$, with terminal node value $\mu_j$) of $T$ to obtain two additional terminal nodes in $T'$ with values $\mu'_j, \mu'_{j+1}$ and set $\mu_j  = \mu'_j = \mu'_{j+1}$. Therefore, with no limitations on the tree topology, the minimum KLD is always zero and, therefore, so is the loss in information. 

Appendix B shows the KLD in the case where there is a maximum number of nodes. In this case, it turns out that the KLD depends on the values at the terminal nodes. Assuming the values at the terminal nodes are i.i.d., then the average KLD over the terminal nodes' distribution is zero. Without averaging over the distribution of the values at the terminal nodes, considering the KLD will affect mostly the prior for the tree with the maximum number of nodes. Intuitively, this is because we can assume that each tree (except the most complex one) is nested into a more complex tree. Given that the outer tree will contain at least the same amount of information as the inner one (as it carries more uncertainty), the loss in information will be zero for any tree except the one with the maximum terminal nodes. For this tree, although the loss in information is not zero, its value is negligible, hence we will consider the loss in information to be always equal zero for the remainder of the paper.

\subsection{Loss in complexity}

We have decided to base the loss in complexity of using tree $T$ on the number of terminal nodes of $T$, namely $n_L(T)$, and the difference between the left and right terminal nodes, $\Delta(T) = |n_l - n_r|$, where $n_l$ ($n_r$) is the number of terminal nodes on the left (right) branch of the tree. The number of terminal nodes is a natural measure of complexity for trees \citep{denison1998bayesian, wu2007bayesian}, while the difference between the number of left and right terminal nodes is included to have control over the skewness of the tree for a given number of terminal nodes. We remark that both the CL and pinball priors penalises for the skewness of the tree, but they do that indirectly given the split probability in the first case, and the distribution of the number of nodes going to the left in the second.

The loss in complexity is then given by

\begin{equation}
Loss_C(T) = -\omega n_L(T) - \gamma \Delta(T),
\label{eq:LC}
\end{equation}
where $\omega \geq 0$ and $\gamma \in \mathbb R$ are weights and will be the parameters of the prior. The loss in complexity is negatively oriented, meaning that less complex models have loss in complexity closer to zero. 

The parameter $\gamma$ is allowed to be negative. Indeed, $\gamma > 0$ means that, given the number of terminal nodes, the prior favours trees with an equal number of terminal nodes on the left and right branches. This leads to shallower trees than when $\gamma < 0$. On the other hand, when $\gamma < 0$, the prior favours trees where the majority of the nodes lies on one of the branches, thus, inducing deeper trees. In this work, we will consider only cases where $\gamma > 0$, which provides a stricter penalty for complexity, because this is the usual case for CART/BART priors; however, the proposed prior is more flexible than the CL prior and works also for $\gamma < 0$, favouring skewed trees, which may be interesting in specific CART applications. 

\subsection{The prior distribution for the BART model}

Considering the expressions given by Equations \ref{eq:LI} and \ref{eq:LC} for the loss in information and complexity, respectively, the LB prior for a tree topology $T$ is given by

\begin{equation}
\pi(T) \propto \exp\{-\omega n_L(T) - \gamma \Delta(T)\}.
\label{eq:priorT_propto}
\end{equation}

In order to find the normalising constant, we consider the following factorisation where, for simplicity, we have dropped the dependence on $T$ of $n_L(T)$ and $\Delta(T)$, 

\begin{equation}
\pi(T) = \frac{\pi(n_L) \pi(\Delta | n_L)}{\mathcal N(n_L, \Delta)},
\label{eq:priorT_equalto}
\end{equation}
where $\mathcal N(n_L, \Delta)$ is the number of binary trees with $n_L$ terminal nodes, and difference between left and right terminal nodes $\Delta$. The analytical expression of $\mathcal N(n_L, \Delta)$ is given in Appendix C. This is equivalent to consider $\pi(n_L)$ for the number of terminal nodes, $\pi(\Delta | n_L)$ for the left and right difference given the number of terminal nodes, and a uniform distribution on the trees with the same $n_L$ and $\Delta$. 

Combining Equations \ref{eq:priorT_propto} and \ref{eq:priorT_equalto}, we have that $\pi(n_L)$ and $\pi(\Delta | n_L)$ are 
\begin{equation*}
\pi(n_L) \propto e^{-\omega n_L}, \quad \text{and}\quad \pi(\Delta|n_L) \propto e^{-\gamma \Delta},
\end{equation*}
from which we can calculate the normalising constants for both distributions.

The calculations for the distribution on the number of terminal nodes (assuming no constraints on the maximum number of terminal nodes) are trivial and give 

\begin{equation*}
\pi(n_L) = \frac{e^{-\omega n_L}}{\sum_{n = 1}^\infty e^{-\omega n}} = e^{-\omega n_L}(e^\omega - 1),
\end{equation*}
which is a Geometric distribution with parameter $p = 1- e^{-\omega}$ (see Appendix D).

The calculations for the conditional distribution of $\Delta$ given $n_L$ are more complicated, and we report them explicitly. The conditional distribution is given by 

\begin{equation*}
\pi(\Delta|n_L) = \frac{e^{-\gamma \Delta}}{\mathcal C(n_L)},
\end{equation*}
where $\mathcal C(n_L)$ is the normalising constant which depends on the value of $n_L$. To calculate $\mathcal C(n_L)$, we start by observing that, if $n_L$ is odd, then $\Delta$ is also odd (being the difference between an odd and an even number) and, symmetrically, if $n_L$ is even, $\Delta$ is even. Furthermore, we consider that $\Delta$ will always be comprised between $0$ and $n_L - 2$ (as we can have at most $n_L - 1$ nodes on one side). This implies that the normalising constant $\mathcal C(n_L)$ for the conditional distribution $\pi(\Delta | n_L)$ must be calculated differently depending on $n_L$ being odd or even, and we need to sum over only the odd or even numbers respectively. Therefore, it is useful to consider a change of variable and use $k = \lfloor \Delta/2 \rfloor$ (where $\lfloor x \rfloor$ is the maximum integer smaller than $x$) instead of $\Delta$ so that 

\begin{equation}
\mathcal C(n_L) = 
\begin{cases}
\sum\limits_{k = 0}^{\lfloor \frac{n_L-2}{2} \rfloor}  e^{-\gamma(2k + 1)} = \frac{e^{\gamma}(1 - e^{-\gamma(n_L - 1)})}{e^{2\gamma} - 1}, & \text{if } n_L \text{ odd}, \\ \\
\sum\limits_{k = 0}^{\lfloor \frac{n_L-2}{2} \rfloor}  e^{-\gamma(2k)} = \frac{e^{2\gamma}(1 - e^{-\gamma n_L})}{e^{2\gamma} - 1}, & \text{if } n_L \text{ even}.
\end{cases}
\label{eq:norm_const_gamma}
\end{equation}

Equation \ref{eq:norm_const_gamma} can be rewritten by considering an indicator function $\delta_o(n_L)$, which is equal to 1 when $n_L$ is odd, and 0 when $n_L$ is even. The equation becomes 

\begin{equation}
\label{eq:norm_const_delta}
C(n_L) =  \frac{\delta_o(n_L)(e^{-\gamma} - 1) + 1 - e^{-\gamma n_L}}{1 - e^{-2\gamma}}.
\end{equation}

From Equation \ref{eq:norm_const_delta} it is clear that the distribution is not proper for $\gamma = 0$ which, indeed, correspond to the uniform case. 

To conclude, the LB prior for the tree $T$ is given by 

\begin{equation*}
\pi(T) = \frac{1}{\mathcal N(n_L, \Delta)}e^{-\omega n_L}(e^\omega - 1) \frac{e^{-\gamma \Delta}}{\mathcal C(n_L)},
\end{equation*}
with $\mathcal C(n_L)$ given by Equation \ref{eq:norm_const_delta}, and $\mathcal N(n_L, \Delta)$, the number of possible trees with given $n_L$ and $\Delta$, given in Equation 4 of Appendix C.

\subsection{Parameter calibration}

In this section, we show the tree depth distribution induced by different parameters of the loss-based prior, and we provide a way to objectively find the values of the parameters to be used as default. As we will show, this is achieved by maximizing the expected loss.

\begin{figure}
\centering
\includegraphics[width=0.9\linewidth]{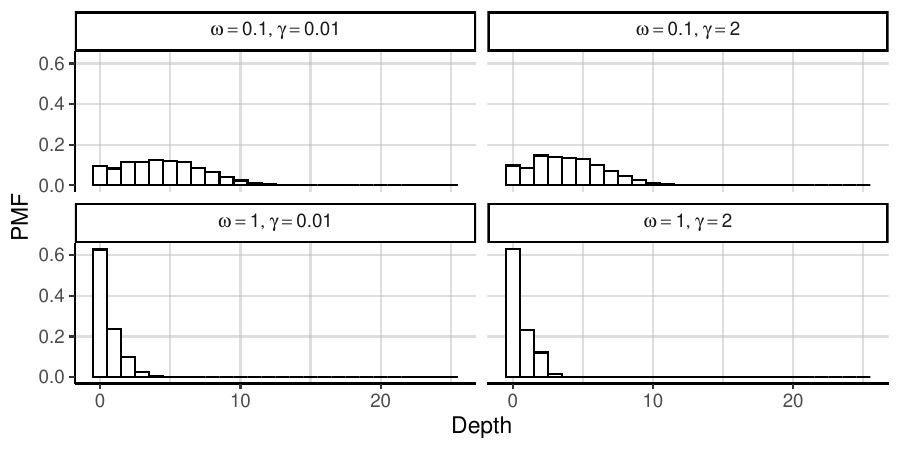}
\caption{Depth probability mass function induced by the LB prior with different values of parameters $\omega$ and $\gamma$, estimated using 10000 tree samples from the prior.}
\label{fig:3_depth_alt}
\end{figure}

\begin{figure}
\centering
\includegraphics[width=0.8\linewidth]{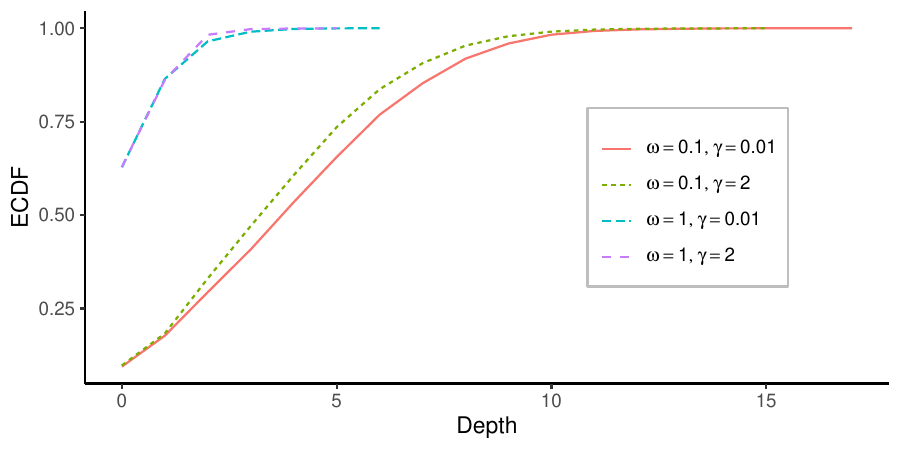}
\caption{Depth empirical cumulative distribution induced by the LB prior with different values of parameters $\omega$ and $\gamma$ (represented by color and line-type), estimated using 10000 tree samples from the prior.}
\label{fig:4_depth_alt_ecdf}
\end{figure}

Figures \ref{fig:3_depth_alt} and \ref{fig:4_depth_alt_ecdf} show the depth probability mass function and cumulative distribution for different values of the parameters $\omega$ and $\gamma$. We do not show the distribution of the number of terminal nodes, which only depends on $\omega$, and is a geometric distribution. Regarding the depth distribution, it is strongly influenced by the value of $\omega$, which determines the expected number of terminal nodes. The effect of parameter $\gamma$ is most appreciable from Figure \ref{fig:4_depth_alt_ecdf}. We can see that $\gamma$ mostly influences the tail of the distribution, and higher values of $\gamma$ provide lighter tails. The case where $\omega$ and $\gamma$ are zero corresponds to the uniform case on the space of possible binary trees. 

\subsubsection{Maximizing the expected Loss}

One way to find a default value for the LB prior parameters is to set an expression for the expected loss, and find the value of the parameters that maximises it. As expected loss, we consider two alternatives providing similar results. These are 

\begin{align*}
E_L^{(1)}(\omega, \gamma) & = \omega^2 \mathbb E(n_L) + \gamma \mathbb E(\Delta), \\
E_L^{(2)}(\omega, \gamma) & = \omega^2 \mathbb E(n_L) + \omega\gamma \mathbb E(\Delta),
\end{align*}
where $\mathbb E(n_L)$ and $\mathbb E(\Delta)$ are the expected values of $n_L$ and $\Delta$ with respect to their marginal distributions. We notice that the marginal expected value of $\Delta$ is a function of both $\omega$ and $\gamma$.

\begin{figure}
\centering
\includegraphics[width=\linewidth]{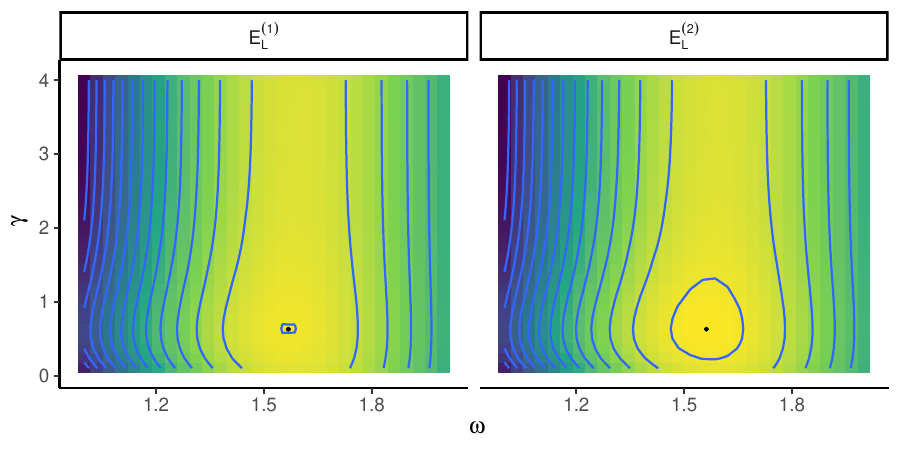}
\caption{Expected loss as a function of LB prior parameters $\omega \in (1,2)$ and $\gamma \in (0,4)$ considering as expected loss the functions $E_L^{(1)}(\omega, \gamma)$ (left) and $E_L^{(2)}(\omega, \gamma)$ (right). In each panel, the black point represents the values of the parameters maximizing the expected loss in each case.}
\label{fig:5_exp_loss}
\end{figure}

Figure \ref{fig:5_exp_loss} shows the expected loss for varying $\omega$ and $\gamma$ using the two above formulations. Under both formulations, we find similar values for the optima, which is $\omega^*_1 = 1.568$ and $\gamma^*_1 = 0.628$ using $E_L^{(1)}(\omega, \gamma)$, and $\omega^*_2 = 1.561$ and $\gamma^*_2 = 0.629$ using $E_L^{(2)}(\omega, \gamma)$. Given this similarity, we proceed considering $E_L^{(2)}(\omega, \gamma)$ as the expected loss function and $\omega^* = \omega^*_2, \gamma^* = \gamma^*_2$ as the default values of the parameters. From now on, when we refer to the default LB prior, we intend the LB prior with parameters $\omega^* = 1.561$ and $\gamma^* = 0.629$.

\section{MCMC for BART}
\label{sec:MCMC}

In this Section, we describe the basic backfitting Markov-Chain Monte Carlo \citep[MCMC,][]{hastie2000bayesian} algorithm used to make inference on CART models introduced by \cite{chipman1998bayesian}. The MCMC algorithm for BART \citep{chipman2010bart} recursively uses the algorithm for CART on each tree, conditional on the value of the others, using the residuals as observations. The process is repeated $m$ times, one for each tree in the summation. The MCMC algorithm for CART is a modification of the Metropolis-Hastings (MH) algorithm \citep{hastings1970monte} where, in each iteration, a new tree is proposed which is then accepted or rejected. For this reason, this section is divided into a first part describing the general MCMC algorithm for BART, and a second part describing the MH algorithm for CART. This is the algorithm we use in Sections \ref{sec:sim_study} and \ref{sec:real_data} to obtain the results. The only difference between the implementations for the CL and LB approaches is how the prior is calculated.

\subsection{Backfitting MCMC for BART}

Here, we describe the MCMC algorithm for BART, assuming the observations follow a Gaussian distribution with unknown variance $\sigma^2$. Given a set of observations, $\mathbf y = (y_1,..,y_n)$, the aim of the MCMC algorithm is used to produce samples from the posterior distribution

\begin{equation*}
\pi((T_1, M_1),...,(T_m, M_m), \sigma | \mathbf y),
\end{equation*}
so that, at each iteration the algorithm produces one posterior sample for the $m$ binary trees $T_1,...,T_m$, for the value at the terminal nodes $M_1,...,M_m$, and for the marginal variance $\sigma$.

The backfitting MCMC algorithm is essentially a Gibbs sampler: each of the $m$ couples $(T_j, M_j)$ is sampled conditionally on the other $m-1$ couples. More formally, we call $\mathbf T_{-j}$ and $\mathbf M_{-j}$ the set of all $m$ trees and terminal nodes values except tree and set $j$. For each MCMC iteration, and $(T_j, M_j), j = 1,...,m$, a sample from the conditional distribution of  
\begin{equation}
(T_j, M_j) | \mathbf T_{-j}, \mathbf M_{-j}, \mathbf y, \sigma,
\label{eq:tree_mcmc}
\end{equation} 
is obtained, followed by a sample from 
\begin{equation}
\sigma | (T_1,M_1),...,(T_m,M_m),\mathbf y.
\label{eq:sigma_mcmc}
\end{equation} 

In order to draw $(T_j, M_J)$ from its conditional distribution, it is important to notice that the conditional distribution in Equation \ref{eq:tree_mcmc} depends on $\mathbf T_{-j}$, $\mathbf M_{-j}$ and $\mathbf y$ only through the vector of residuals $\mathbf R_j = (R_{1j},...,R_{nj})$, where the single component is given by 

\begin{equation}
R_{ij} = \mathbf y_i - \sum_{k \neq j} g(\mathbf x_i, T_k, M_k) .
\label{eq:residual_tree}
\end{equation}

So, extracting a sample from Equation \ref{eq:tree_mcmc} is equivalent to sampling the posterior of a single tree model 

\begin{equation*}
(T_j, M_j)| \mathbf R_j, \sigma,
\end{equation*}
where $\mathbf R_j$ plays the role of the observations. 

In this context, if a conjugate prior is used for the terminal node values, the posterior for the tree is given by 

\begin{equation*}
\pi(T_j | \mathbf R_j, \sigma) = \pi(T_j)\int \pi(\mathbf R_j|T_j,M_j,\sigma)\pi(M_j|T_j,\sigma)\,dM_j,
\end{equation*}
which is available in close form, and we can draw a sample from the posterior distribution of $(T_j,M_J)|\mathbf R_j, \sigma$ by sampling a tree from the posterior of 

\begin{equation*}
T_j | \mathbf R_j, \sigma.
\end{equation*}

This step is  performed using the Metropolis-Hastings MCMC algorithm for CART models using $\mathbf R_j$ as observations. Then, conditionally on the obtained tree, we sample the value at the terminal nodes from   
\begin{equation*}
M_j|T_j,\mathbf R_j, \sigma.
\end{equation*}

It is common practice to assume a conjugate prior for the values at the terminal nodes, and therefore, for this step we can sample directly from the conditional posterior distribution. For example, considering the BART model in Equation \ref{eq:BART_def}, and assuming a normal prior for the terminal node values, also the posterior distribution is normal.

Lastly, we assume a conjugate prior distribution for the marginal variance. For example, considering the BART model in Equation \ref{eq:BART_def}, if an Inverse-Gamma distribution is chosen as prior for the marginal variance, the posterior of the marginal variance conditional on the values of $(T_1,M_1),...,(T_m,M_m)$ it is also an Inverse-Gamma distribution, and therefore we can sample directly from it. 

The algorithm for BART is summarised in steps in Algorithm  \ref{alg:mcmc_bart}.

\begin{algorithm}
  \caption{MCMC for BART - $(i+1)$-th iteration}
  \label{alg:mcmc_bart}
  \begin{algorithmic}[1]
    \State Previous step produces $(T_j^{(i)},M_j^{(i)})$ for $j=1,...,m$, and $\sigma^{(i)}$ 
     \For{j = 1,...,m}
     	\State Calculate residuals $\mathbf R_{j}^{(i)}$ using Equation \ref{eq:residual_tree}
        \State Sample $T_j^{(i+1)}$ from $T_j|\mathbf R_{-j}^{(i)}, \sigma$ \Comment{Using algorithm for CART}
        \State Sample $M_j^{(i+1)}$ from $M_j|T_j^{i+1},\mathbf R_j, \sigma$\Comment{From the conjugate posterior}
      \EndFor
      \State Sample $\sigma^{(i+1)}$ from $\sigma|(T_1^{(i+1)},M_1^{(i+1)}),...,(T_m^{(i+1)},M_m^{(i+1)}), \mathbf y$\Comment{From the conjugate posterior}
  \end{algorithmic}
\end{algorithm}

\subsection{Metropolis-Hastings MCMC for CART}

In this Section, we describe the Metropolis-Hastings MCMC algorithm \citep{chipman1998bayesian} used to sample $(T,M)$ from the conditional distribution of $T|\mathbf y,\sigma$. This algorithm was developed to explore the posterior distribution of the tree of a CART model, which is an instance of the BART model described by Equation \ref{eq:BART_def}, considering $m = 1$. The goal is to be able to generate samples $T^{(0)},T^{(1)},T^{(2)},\dots$ from the posterior distribution $\pi(T|\mathbf y, \sigma)$. 

At each iteration, the algorithm is based on proposing a new candidate tree $T^*$, and accepting/rejecting it based on some probability. At iteration $i+1$, the new tree is proposed by performing a \emph{move} on the previous tree $T^{(i)}$. Here, we describe the original algorithm in which a new tree is proposed according to one of four possible moves (GROW, PRUNE, SWAP, CHANGE). However, we notice that much research has been done in designing alternative moves providing better mixing \citep{wu2007bayesian, pratola2016}, but this is beyond the scope of this article.

To generate the tree $T^*$ from $T^{(i)}$, we pick at random one of the following possible moves

\begin{itemize}
\item GROW: Randomly choose a terminal node and split it into two additional terminal nodes. The new splitting rule is assigned according to the prior.
\item PRUNE: Randomly choose a parent of a terminal node and turn it into a terminal node by removing its children.
\item SWAP: Randomly choose a parent-child pair of internal nodes and swap their splitting rules.
\item CHANGE: Randomly choose an internal node and assign a new splitting rule according to the prior distribution.
\end{itemize} 

The moves are performed to ensure that $T^*$ yields a \emph{valid} partition, as per Definition 2 in Appendix A, for some $C^2 \geq 1$. These moves define a transition kernel $q(T, T^*)$ given by the probability of obtaining $T^*$ from $T$. An appealing feature of this kernel is that it produces a reversible Markov chain given that the PRUNE and GROW moves are counterparts, as well as SWAP and CHANGE.

Once a tree $T^*$ is produced from tree $T^{(i)}$ according to $q(T^{(i)}, T^*)$, it is accepted with probability

\begin{equation}
\alpha(T^{(i)}, T^*) = \min\left(\frac{q(T^*, T^{(i)})\pi(\mathbf y| T^*)\pi(T^*)}{q(T^{(i)}, T^*)\pi(\mathbf y| T^{(i)})\pi(T^{(i)})}, 1\right).
\label{eq:trans_prob}
\end{equation} 
Otherwise, the tree does not change.

To summarise, the algorithm to explore the tree posterior of a CART model is given in Algorithm \ref{alg:mcmc_cart}.

\begin{algorithm}
  \caption{MCMC for CART - $(i+1)$-th iteration}
  \label{alg:mcmc_cart}
  \begin{algorithmic}[1]
    \State The previous step produces $T^{(i)}$ 
    \State Generate $T^* \sim q(T^{(i)}, T^*)$
    \State Compute acceptance probability $\alpha(T^{(i)}, T^*)$ according to Equation \ref{eq:trans_prob}
    \State Set $T^{(i+1)} = T^*$ with probability $\alpha(T^{(i)}, T^*)$, or $T^{(i+1)} = T^{(i)}$ otherwise
  \end{algorithmic}
\end{algorithm}

\section{Simulation Study}
\label{sec:sim_study}

In this section, we simulate observations from a known CART model with Gaussian residuals and compare the posterior distributions obtained using different instances of the LB and CL priors. The goal is to show that the LB prior provides a greater penalty for complexity than the CL (assuming similar prior expectations), without providing worse performances in terms of mean square error in out-of-sample prediction experiments. We also explore how this affects the frequency with which predictors are used. The choice of using a CART model, which is a case of BART model with $m=1$, is motivated by the fact that the MCMC algorithm described in the previous section is influenced by the prior only when computing the acceptance probabilities in the MCMC for CART, and therefore it represents a natural first step. Furthermore, given that the MCMC algorithm for BART is a recursive application of the CART algorithm, the effects of using the LB prior that we see in this example will translate to the BART case, as shown in Section \ref{sec:diabetes}. 

We fit six different CART models using the MCMC algorithm for CART described in Section \ref{sec:MCMC} corresponding to three LB priors and three CL priors with different parameters. The parametrisations are such that they match the expected number of terminal nodes and depth between LB and CL priors (see Table \ref{tab:1_prior_sim}). We consider the marginal variance $\sigma^2$ as known, while for the splitting rules we consider a prior distribution given by assuming a discrete uniform distribution on the space of available predictors for the splitting variable, and a continuous uniform distribution between a range of available values for the splitting value. 

For the data, we simulate predictors and observations with the same model used by \cite{wu2007bayesian} in their simulation experiment on the pinball prior. We consider 300 observations according to a single tree model ($m = 1$) considering 3 available predictors $X_1, X_2, X_3$ distributed according to

\begin{align*}
X_{1j} & \sim 
\begin{cases}
\text{Unif}(0.1,0.4) & j = 1,\dots,200\\ 
\text{Unif}(0.6,0.9) & j = 201,\dots,300
\end{cases} \nonumber \\
X_{2j} & \sim 
\begin{cases}
\text{Unif}(0.1,0.4) & j = 1,\dots,100\\ 
\text{Unif}(0.6,0.9) & j = 101,\dots,200 \\
\text{Unif}(0.1,0.9) & j = 201,\dots,300
\end{cases} \\ 
X_{3j} & \sim 
\begin{cases}
\text{Unif}(0.6,0.9) & j = 1,\dots,200\\ 
\text{Unif}(0.1,0.4) & j = 201,\dots,300,
\end{cases} \nonumber
\end{align*}
and observations given by 
\begin{equation}
Y_j = 
\begin{cases}
1 + N(0,0.25) & \text{if } X_{1j} \leq 0.5, X_{2j} \leq 0.5 \\
3 + N(0,0.25) & \text{if } X_{1j} \leq 0.5, X_{2j} > 0.5 \\
5 + N(0,0.25) & \text{if } X_{1j} > 0.5.
\end{cases}
\label{eq:sim_model}
\end{equation}

\begin{figure}
\centering
\includegraphics[width=\linewidth]{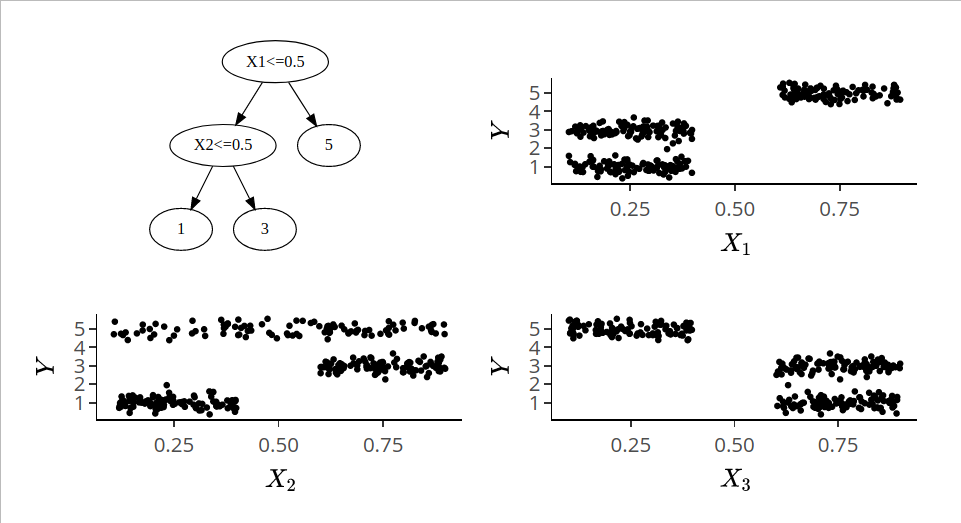}
\caption{Tree and scatter plots of 300 observations taken from the model specified in Equation \ref{eq:sim_model}.}
\label{fig:6_sim_data}
\end{figure}

We assume the model described by Equation \ref{eq:sim_model} is the same as the model in Equation \ref{eq:BART_def} with $m=1$, and considers one regression tree $T$, given by the top-left panel of Figure \ref{fig:6_sim_data}. The marginal variance is $\sigma^2 = 0.25$. In this model, only predictors $X_1$ and $X_2$ are actually used to generate $Y$, while $X_3$ is there as disturbance term. Figure \ref{fig:6_sim_data} shows scatter plots of $Y$ against $X_1, X_2, X_3$.

We consider six priors (Table \ref{tab:1_prior_sim}): three instances of the CL prior and three of the LB prior assuming different parameters values. We include the two default priors (CL: $\alpha = 0.95, \beta = 2$, LB: $\omega = 1.56, \gamma = 0.62$), then two LB priors matching the default CL prior expected number of terminal nodes ($2.51$) and depth ($1.45$), and two CL priors matching the default LB prior expected number of terminal nodes ($1.26$), and depth ($0.25$).

\begin{table}
\begin{center}
\resizebox{\columnwidth}{!}{%
\begin{tabular}{ llrrrrrrrr}
 \hline
 Prior & Parameters & $\mathbb E(n_L)$ & $\Pr(n_l > 9)$ & 
 $\mathbb E(D)$ & $\Pr(D > 6)$ & $\mathbb E(n_L|Y)$ & $\Pr(n_l > 9 | Y)$& $\mathbb E(D|Y)$ & $\Pr(D >6 | Y)$ \\
 \hline 
 CL & $\alpha = 0.95, \beta = 2$ & 2.51 & $<10^{-3}$ & 1.45 & $<10{-3}$ & 8.399 & 0.346 & 4.66 & 0.166\\ 
 CL & $\alpha = 0.25, \beta = 2$ & 1.29 & $<10^{-3}$ & 0.28 & $<10^{-3}$ & 7.241 & 0.238 & 4.17 & 0.106\\ 
 CL & $\alpha = 0.25, \beta = 0.5$ & 1.36 &$<10^{-3}$ & 0.35 & $<10^{-3}$ & 7.494 & 0.280 & 4.16 & 0.126\\
 \hline
 LB & $\omega = 1.56, \gamma = 0.62$ & 1.26 & $<10^{-3}$ & 0.25 & $<10^{-3}$ & 5.946 & 0.123 & 3.40 & 0.046\\ 
 LB & $\omega = 0.5, \gamma = 0.62$ & 2.52 & 0.03 & 1.21 & $<10^{-3}$ & 8.077 & 0.317 & 4.49 & 0.153\\ 
 LB & $\omega = 0.5, \gamma = 1.5$ & 2.52 & 0.03 & 1.2 & $<10^{-3}$ & 8.004 & 0.305 & 4.33 & 0.098\\
 \hline
\end{tabular}
}
\end{center}
\caption{\label{tab:1_prior_sim} Priors compared in the simulation experiment. The columns represents the type of prior (CL: classic prior,  LB: loss-based prior), the prior expected number of terminal nodes, the prior probability that the number of terminal nodes is greater than 9, the prior expected depth, the prior probability that the depth is greater than 6, the posterior expected number of terminal nodes, the posterior probability that the number of terminal nodes is greater than 9, the posterior expected depth, and the posterior probability that the depth is greater than 6. Prior quantities are estimated using a sample of 10000 trees from the priors, while posterior quantities are obtained by running 250 chains, with 750 samples per chain, and considering a burn-in of 250 samples per chain.}
\end{table}

For each model we run the MCMC algorithm described in Section \ref{sec:MCMC} considering the marginal variance $\sigma^2 = 0.25$ to be known (as we are only interested in the effect of different priors for the tree topology). All the models assume the same prior on the splitting rules, and the values at the terminal nodes. We run 500 chains in parallel and each chain comprises 500 MCMC samples with 250 burn-in. The trace plots (shown in Appendix F) highlight the usual behavior described in \cite{chipman1998bayesian}; each chain quickly converges to a high likelihood region and stays there, exhibiting poor mixing. 

\begin{figure}[H]
\centering
\includegraphics[width=0.9\linewidth]{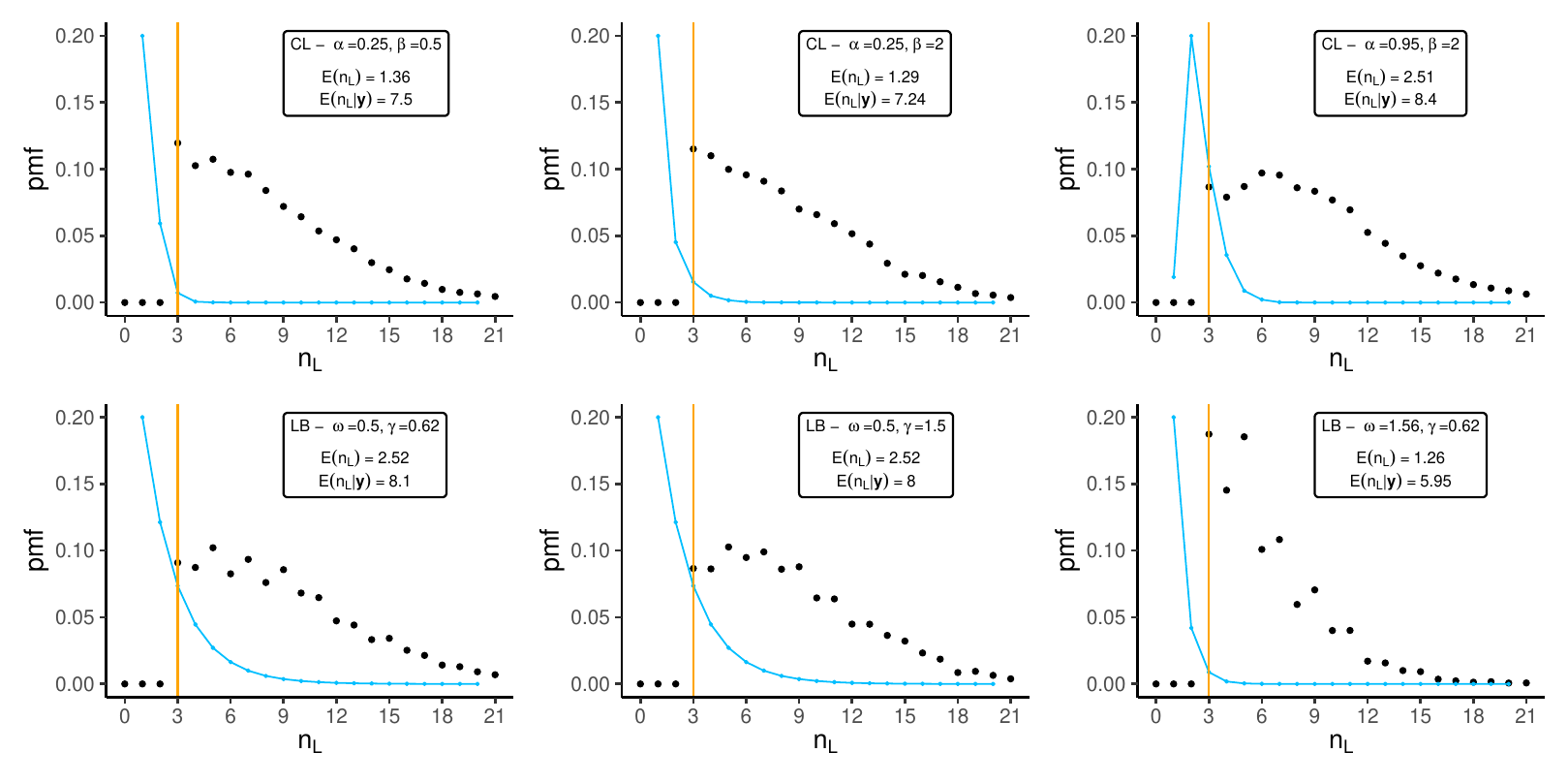}
\caption{Number of terminal nodes posterior distributions (black dots) using different prior distributions for the tree topology. The posteriors are obtained by running 500 parallel chains each one composed of 500 iterations and burn-in of 250. The priors considered are the classic tree prior (CL) with parameter couples $(\alpha, \beta) = \{(0.25,0.5), (0.25,2), (0.95,2)\}$ and, the LB prior with parameters couples $(\omega,\gamma) = \{(0.5,0.62),(0.5,1.5),(1.56,0.62)\}$. The light blue lines represent the prior distributions, while the vertical orange line represents the true number of terminal nodes. Priors have been scaled to have their maximum at $0.2$.}
\label{fig:post_nl}
\end{figure}

Figures \ref{fig:post_nl} shows the posterior distribution of the number of terminal nodes for the different priors listed in Table \ref{tab:1_prior_sim}. Comparing the default CL prior ($\alpha = 0.95, \beta = 2$) against LB priors with similar expectations ($\omega = 0.5, \gamma = 0.62, 1.5$) there is no much difference. The LB prior only provides slightly shorter trees with a smaller probability of exceedance (see Table \ref{tab:1_prior_sim}). It is interesting to notice that the LB approach provides posterior distributions concentrated around smaller values of $n_L$ than using the CL prior, even though the exceedance prior probability ($\Pr(n_L > 9)$) is smaller for the CL prior. Comparing the LB default prior ($\omega = 1.56, \gamma = 0.62)$ with CL priors with similar expectations ($\alpha = 0.25, \beta = 0.5, 2$) we observe a more pronounced difference between the posterior distributions, with the LB prior being concentrated around smaller values. This shows that considering priors with similar expectations the LB approach provides a greater penalty for increasing the complexity of the tree than the CL approach. We can observe the same effect on the depth distribution in Figure 1 of Appendix E.

\begin{figure}[H]
\centering
\vspace{-1cm}
\includegraphics[width=0.9\linewidth]{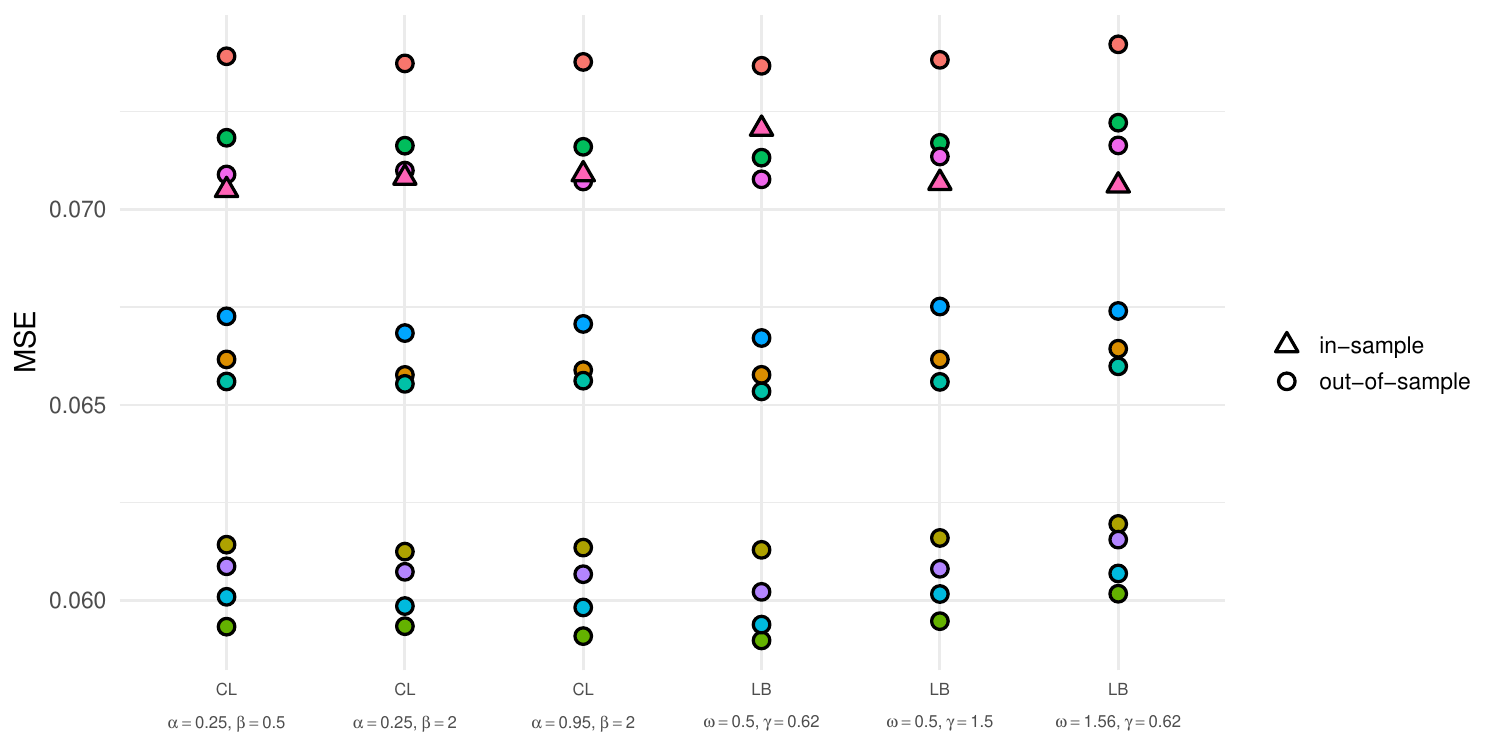} 
\includegraphics[width=0.9\linewidth]{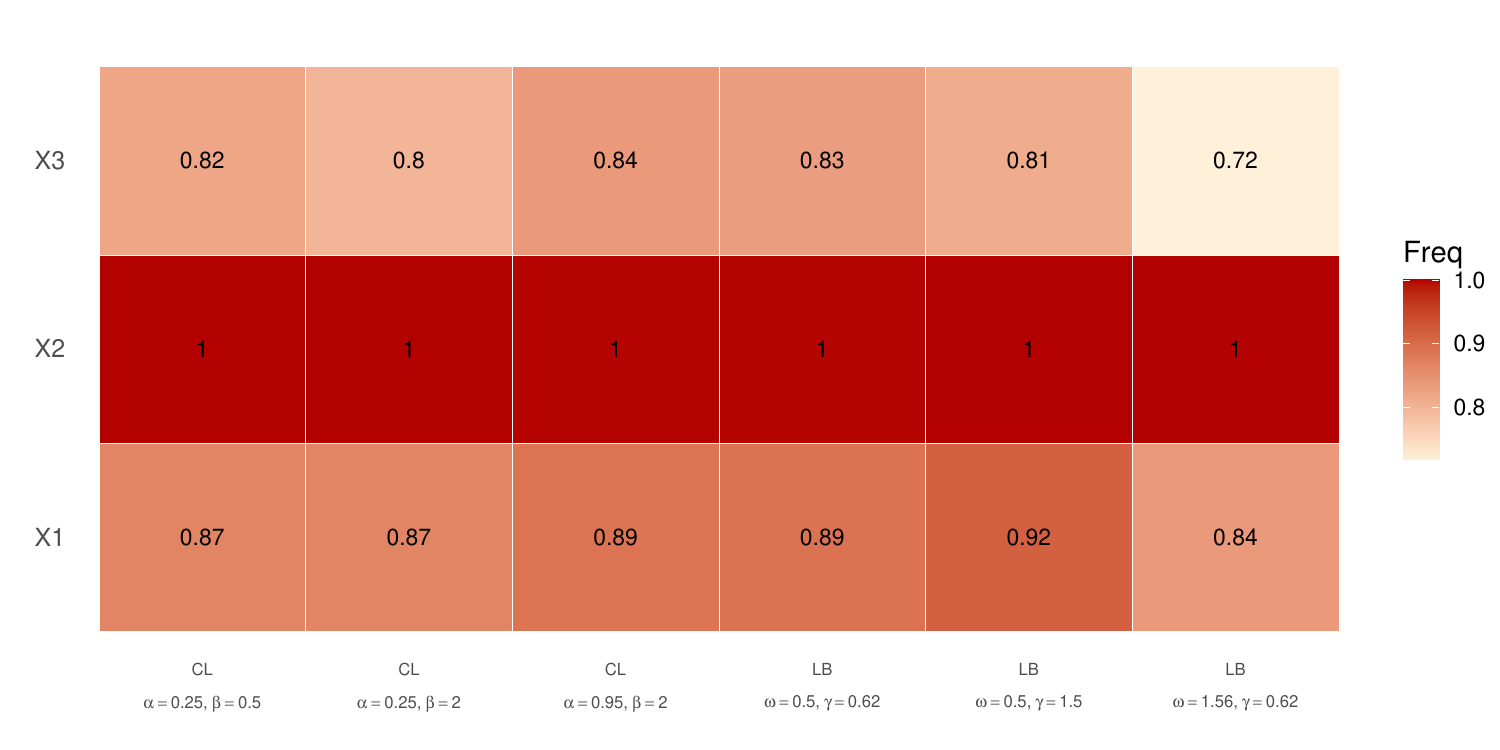}
\caption{Top: Mean Square Error (MSE) for different prior distributions on 10 different out-of-sample prediction experiments (circles) and in-sample (triangles). Bottom: Percentage with which each predictor is selected in trees explored by the MCMC algorithm for different prior distributions. The priors considered are the classic tree prior (CL) with parameters couples $(\alpha, \beta) = \{(0.25,0.5),(0.25,2),(0.95,2)\}$ and, the LB prior with parameter couples $(\omega,\gamma) = \{(0.5,0.62),(0.5,1.5),(1.56,0.62)\}$.}
\vspace{-1cm}
\label{fig:mse_synth}
\end{figure}

The fact that the LB prior explores shorter trees than the CL prior may lead to the model performing worse. To check this, we considered 10 out-of-sample prediction experiments. In each one, we simulate 300 new predictors' values and observations according to the model in Equation \ref{eq:sim_model}, calculate the predictions under each model (we use the posterior mean as estimator), and compute the Mean Square Error (MSE). Results for the different out-of-sample experiments are shown in Figure \ref{fig:mse_synth} (top). All the models considered performed very similarly, and the MSE of the out-of-sample experiments is in line with the in-sample MSE, which shows that the models did not overfit the data. This shows that the LB prior has similar performance to the CL prior, but using shorter trees. This is relevant both computationally, it takes less time to compute the function $g(\bx, T, M)$ for shorter trees, but also for the interpretation. Indeed, considering shorter trees stimulates competition between predictors to enter the pool, ultimately yielding a more refined and informative predictor set. This limits the tendency of the model to explore unnecessarily complex trees, but pushes it instead to optimise shorter ones. This is confirmed by looking at trees with the highest likelihood explored during the MCMC routine (Figure 6 of Appendix G of the supplementary material) where the LB default prior is the one providing the shorter tree.

To show the effect of having shorter trees on the predictors' selection, we calculated the fraction of trees explored during the MCMC in which each predictor is selected. The results are shown in Figure \ref{fig:mse_synth} (bottom). We see that all models provides a consistent hierarchy: $X_2$ is always selected, followed by $X_1$, followed by $X_3$ which was the disturbance term. However, we notice that the LB prior is the one providing the greater separation in the times each predictor is selected. Specifically, the model that selects $X_3$ the least is the default LB prior ($\omega = 1.52, \gamma = 0.62$) showing the effect of considering shorter trees.

To conclude, in this simple synthetic CART experiment we have shown that LB prior provides a greater penalty for increasing complexity than the CL prior, considering the two having similar prior expectations on the number of terminal nodes and depth. This makes the MCMC algorithm explore shorter trees under the LB prior without \emph{loosing} in out-of-sample forecasting capabilities. In turns, this increases competition among predictors providing better variable selection. In the next Section, we show that these effects holds when using real data, and can be generalised to the BART case.

\section{Real Data applications}
\label{sec:real_data}

In this section, we provide two real data applications: one using CART, and one using the BART. In both, we compare the results using the LB priors introduced in this article, and the CL priors described in \cite{chipman1998bayesian}. The first application uses CART, and is on the breast cancer data \citep{wolberg1990multisurface} which was already analysed by different authors \citep{breiman1996bagging, chipman1998bayesian, wu2007bayesian}, and constitutes a benchmark application for the CART model. The second application is on diabetes data \citep{diabetes} for which the BART model has never been used before (according to the authors knowledge). The main aim of this section is to show that the advantages of using the LB prior on CART models reported in Section \ref{sec:sim_study} using synthetic data also hold when analysing real data, and for BART models.

\subsection{Breast cancer data}
\label{sec:breast_cancer}
In this first real data application, we apply the CART model to the breast cancer data collected by William HI. Wolberg, University of Wisconsin Hospitals, Madison \citep{wolberg1990multisurface}, \citep{breiman1996bagging, chipman1998bayesian, wu2007bayesian}. The dataset can be downloaded from the University of California Irvine repository of machine-learning databases \footnote{\url{https://archive.ics.uci.edu/dataset/15/breast+cancer+wisconsin+original}}. The variable of interest  is a binary variable indicating whether cancer is malignant (1) or benign (0), therefore we need to consider a different likelihood for the observations than the one provided in Equation \ref{eq:BART_def}. We focus on comparing the default LB prior ($\omega = 1.56, \gamma = 0.62$), and the default CL prior ($\alpha = 0.95, \beta = 2$). Results for additional parametrisations (CL $\alpha = 0.95, \beta = 1,1.5$, LB $\omega = 0.3, 0.42, \gamma = 0.5, 1.5$) are reported in Appendix H of the supplementary material. The aim is to show that under the LB prior we visit shorter trees without loosing in terms of performance which here is measured by the missing rate. As in the previous Section, we report how many times each predictor is selected to highlight the most important ones. Below, we first introduce the data and the likelihood, and then explore the results.

\subsubsection{Data and Likelihood}  

The data provides 9 different predictors corresponding to different cellular characteristics listed in Table \ref {tab:2_breast_cancer_vars} and 699 observations. There are 16 observations with missing Bare Nuclei value, we discard them and use only 683 observations in the analysis. The predictors are normalised to vary between 0 and 1.  

\begin{table}[]
\begin{center}
\resizebox{0.6\columnwidth}{!}{%
\begin{tabular}{ lrrr}
 \hline
 Variable & Range & Encoding & $Cor(X_i,Y)$ \\
 \hline
 \hline
 Clump Thickness & 1-10 & $X_1$ & 0.714\\
 Uniformity of Cell Size & 1-10 & $X_2$ & 0.820\\
 Uniformity of Cell Shape & 1-10 & $X_3$ & 0.821\\
 Marginal Adhesion & 1-10 & $X_4$ & 0.706\\
 Single Epithelial Cell Size & 1-10 & $X_5$ & 0.690\\
 Bare Nuclei & 1-10 & $X_6$ & 0.822\\
 Bland Chromatin & 1-10 & $X_7$ & 0.758\\
 Normal Nucleoli & 1-10 & $X_8$ & 0.718\\
 Mitoses & 1-10 & $X_9$  & 0.423 \\
 \hline
\end{tabular}
}
\end{center}
\caption{\label{tab:2_breast_cancer_vars} Table of available predictors in the breast cancer dataset. First column represents the name of the variable, the second represents the domain of the variable, the third represents how the variable is encoded, the fourth column reports the correlation with $Y$ assuming the value 1 if the cancer is malign and 0 otherwise.}
\end{table}

The observations are binary in this case and therefore we consider a Bernoulli likelihood with probability given by the value of the tree. The probability is modelled with a CART model, so using a single regression tree. More formally, given a set of $n$ observations $\mathbf Y \in \{0,1\}^n$ ($n = 683$), a predictors' matrix $\mathbf X \in [0,1]^{n\times p}$ where each row represents an observation with $p$ predictors ($p = 9$), and a tree $T$ with terminal nodes values $M$, we consider

\begin{equation}
\pi(\mathbf Y | \mathbf X, T, M) = \prod_{i = 1}^n p_i^{Y_i}(1-p_i)^{1 - Y_i}, 
\label{eq:binary_lik}
\end{equation}
where $\bx_i$ is the $i$-th row of $\mathbf X$, $p_i$ is the probability of observing $Y_i = 1$, and it is given by the value of the terminal node corresponding to predictor value $\bx_i$, namely $p_i = g(\bx_i, T, M)$.

Given the new likelihood and role of the terminal nodes values $M = (\mu_1,...,\mu_{n_L(T)})$, which have to satisfy $\mu_j \in [0,1]$, the corresponding conjugate prior for $\mu_j$ is a Beta distribution with parameters $\alpha_\mu, \beta_\mu$. The conditional posterior distribution is again a Beta distribution with parameters 

\begin{equation*}
\mu_j | \mathbf Y, \mathbf X, T \sim \text{Beta} \left(\alpha_\mu + \sum_i Y_{ij},  \beta_\mu + n - \sum_i Y_{ij} \right),
\end{equation*}
where $Y_{ij}$ are the observations associated with the $j$-th terminal node. 
In this example, we consider $\alpha_\mu = 1, \beta_\mu = 1$ which corresponds to the uniform prior on the $[0,1]$ interval.

\subsubsection{Results}

\begin{figure}
\centering
\includegraphics[width=0.9\linewidth]{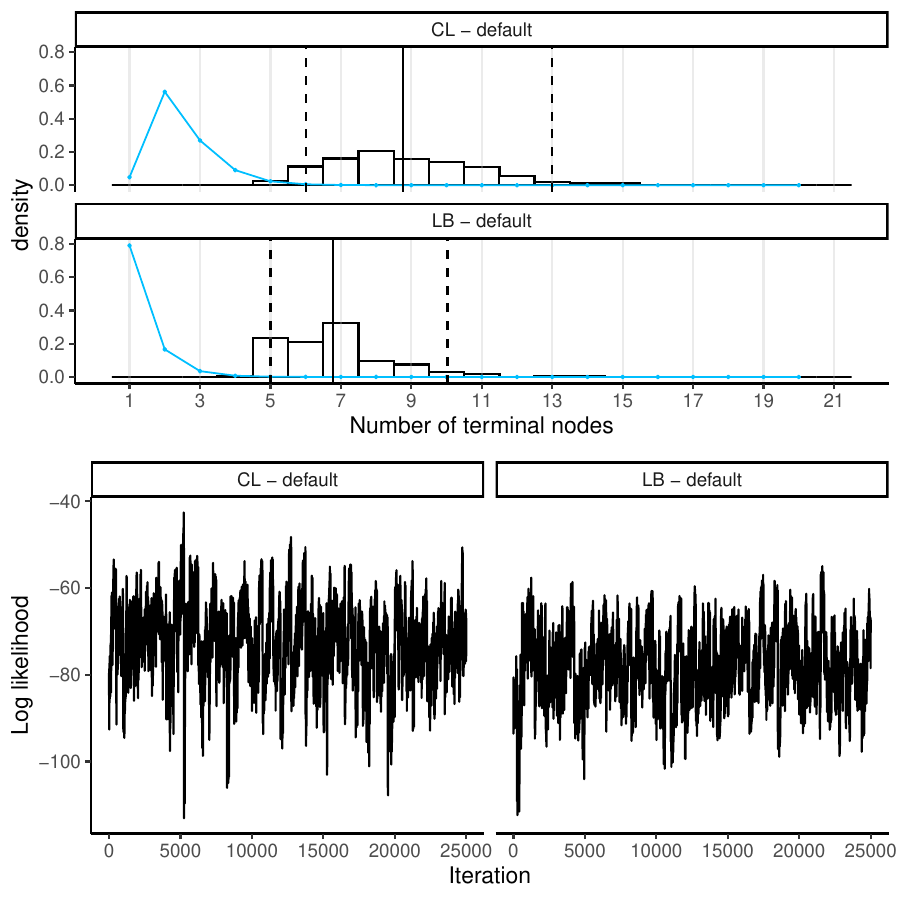}
\caption{Top panels: Breast cancer data number of terminal nodes posterior distributions using the classic prior with default parameters ($\alpha = 0.95, \beta = 2$, first top panel) and the LB prior with default parameters ($\omega  = 1.56, \gamma = 0.62$, second top panel). Light blue lines represent the prior distributions, black solid vertical lines represent the posterior means while the dashed lines represent $95\%$ posterior credibility intervals. Posterior results are obtained from 100 independent chains each composed of 500 posterior samples and considering a burn-in of 250 samples. Bottom panels: log-likelihood traceplots of the selected posterior samples. }
\label{fig:bc_post_nterm}
\end{figure}

In this section, we compare the posterior results obtained using the default CL prior with parameters $\alpha = 0.95, \beta = 2$ and the default LB prior with $\omega = 1.56, \gamma = 0.62$. The two methods yield prior distributions for the number of terminal nodes and depth with different means and tail probabilities, as reported in Table \ref{tab:1_prior_sim} (light blue rows). Therefore, we expect the default LB prior to provide a stronger penalty and, consequently, to explore shorter trees during the MCMC routine. For both cases, we run 100 chains with starting tree the trivial tree with only one node. Figure \ref{fig:bc_post_nterm} shows the traceplots of the log-likelihood for the posterior samples used in the analysis, from which it appears that the chains converge. We also considered the CL prior with parameters $\alpha = 0.95, \beta = 1, 1.5$, which performed well in \cite{chipman1998bayesian}, and the LB prior with parameters $\omega = 0.3, 0.42, \gamma = 1.5, 0.5$, replicating the expected number of terminal nodes as the CL prior; results for this additional cases are shown in Appendix H as they do not provide further insights.

\begin{figure}
\centering
\includegraphics[width=0.9\linewidth]{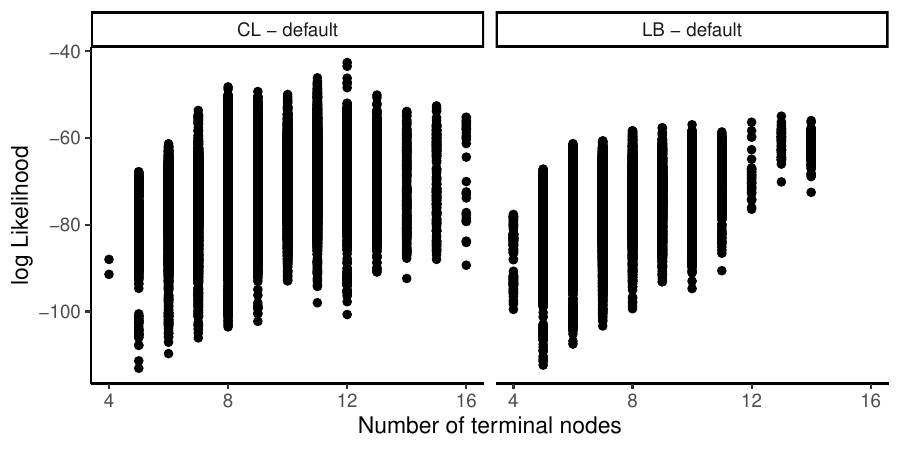}
\caption{Log likelihood as a function of the number of terminal nodes of the trees explored during the MCMC routine.}
\label{fig:bc_post_nterm_loglik}
\end{figure}

\begin{figure}
\centering
\includegraphics[width=0.9\linewidth]{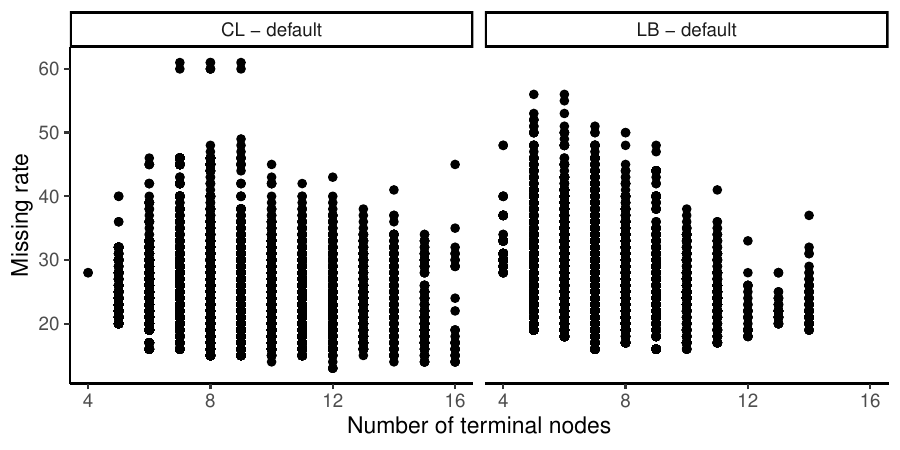}
\caption{Missing rate as a function of the number of terminal nodes of the trees explored during the MCMC routine.}
\label{fig:bc_post_nterm_missrate}
\end{figure}

Looking at the posterior distributions in Figure \ref{fig:bc_post_nterm}, we can see that, as expected, the default LB prior provides a posterior distribution concentrated around lower values of the number of terminal nodes. Indeed, the posterior mean of the number of terminal nodes (vertical solid lines) are 8.75 for the CL prior, and 6.76 for the LB tree prior, while the $95\%$ posterior credibility intervals (vertical dashed lines) are $(6,13)$ for the CL prior, and $(5,10)$ for the LB prior. We notice that the trees explored by the LB prior, despite being shorter, provide the same levels of log-likelihood as those explored by the CL prior. This is confirmed by Figure \ref{fig:bc_post_nterm_loglik}, which shows that using the LB prior all the trees visited with more than 12 nodes have high log-likelihood, while when using the CL prior they are more disperse. The same is true if we look at the missing rates versus the number of terminal nodes shown in Figure \ref{fig:bc_post_nterm_missrate}.

\begin{figure}
\centering
\includegraphics[width=0.9\linewidth]{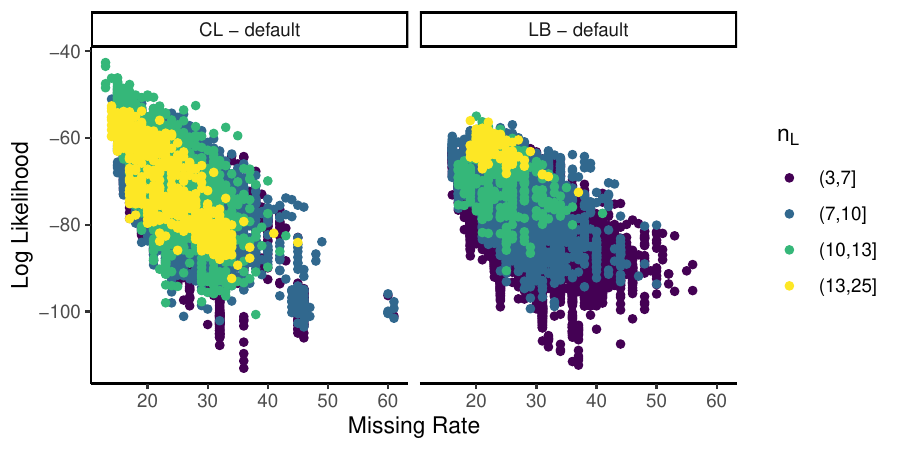}
\caption{Log likelihood as a function of the missing rate for different number of terminal nodes classes (color) of the trees explored during the MCMC routine.}
\label{fig:bc_post_nterm_loglik_missrate}
\end{figure}

Figure \ref{fig:bc_post_nterm_loglik_missrate} combines the information in Figures \ref{fig:bc_post_nterm_loglik} and \ref{fig:bc_post_nterm_missrate}. From this figure is clear that using the LB prior the trees with high numbers of terminal nodes are also the ones with the highest log-likelihood and less variable missing rate. In contrast, using the CL prior with default parameters, the trees with high number of terminal nodes do not provide a clear advantage in terms of log-likelihood and missing rate than the simpler ones. Looking at the LB prior we see that ,as the number of terminal nodes increases, the log-likelihood is more concentrated around high values; this is not as clear for the CL prior., which is an important benefit of using the default LB prior over the default classic prior for this problem, as it provides shorter trees with the same predictive capabilities and log-likelihood than the more complex ones explored under the CL prior. Overall, the CL prior is capable of achieving the highest log-likelihood (CL$ = -42.60$, LB$ = -54.97$) but this does not translate into a sensible difference in the missing rate (CL$ = 13$, LB$ = 16$). This is to show that for this type of model, it is not always the model with the highest likelihood that is the one providing better forecasts. Indeed, for the LB case, the minimum missing rate is achieved by a tree with log-likelihood around $-62$, while the one with highest likelihood has a missing rate of $20$.

\begin{figure}
\centering
\includegraphics[width=0.9\linewidth]{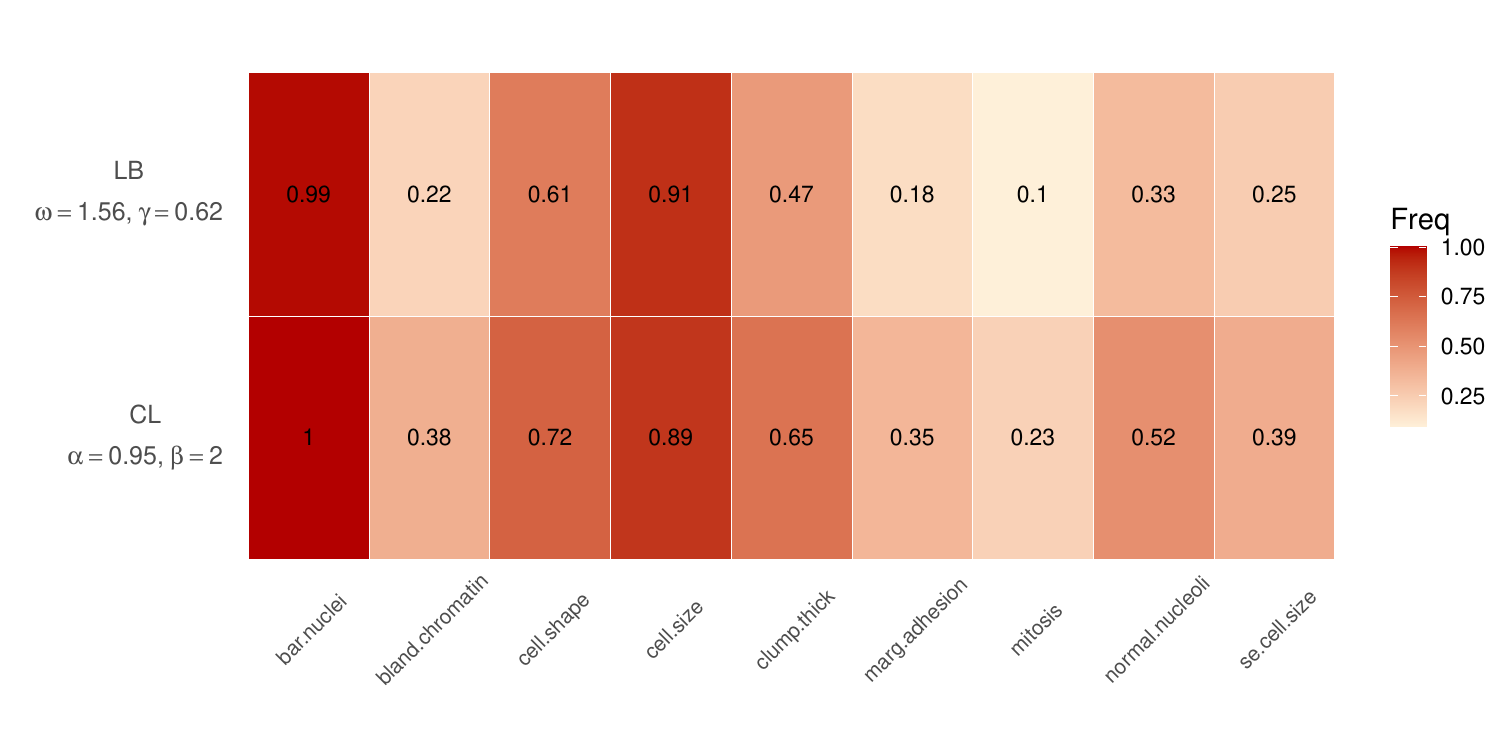}
\caption{Breast cancer data: percentage with which each predictor is selected in trees explored by the MCMC algorithm for the CL default prior $(\alpha = 0.95, \beta = 2)$, and the LB default prior $(\omega = 1.56, \gamma = 0.62)$.}
\label{fig:bc_pred_def}
\end{figure}

We also analysed the frequency with which each predictor is used by a tree explored during the MCMC routine. Figure \ref{fig:bc_pred_def} confirms what we have shown during the synthetic examples. Exploring shorter trees, increases competition among predictors, which provides a more concise set of relevant predictors. Indeed, under the LB prior, important predictors as \emph{bare nuclei}, and \emph{cell size} are selected with the same frequency as under the CL prior. However, non-relevant predictors as \emph{mitosis} and \emph{marginal adhesion}, with participation percentage 0.1 and 0.18 under the LB prior, have percentage  0.23 and 0.35 under the CL prior.

\subsection{Diabetes data}
\label{sec:diabetes}

In this Section, we compare the LB prior and the CL prior using the BART data. We use a dataset \citep{diabetes} on diabetes patients comprehending ten years (1999-2008) of observations from 130 US hospitals and integrated delivery networks. Each entry represents a diabetic patient who underwent laboratory, medications, and stayed up to 14 days. As for the breast cancer data, the variable of interest is binary and indicates whether early readmission of the patient within 30 days of discharge. The dataset can be downloaded from the UC Irvine machine learning repository \footnote{\url{http://archive.ics.uci.edu/dataset/296/diabetes+130-us+hospitals+for+years+1999-2008}}. As priors, we considered the LB default setting $\omega = 1.62, \gamma = 0.62$, while for the CL prior, we consider $\alpha = 0.25, \beta = 2$ already used during the synthetic experiment in Section 5. Both priors provide a distribution for the number of terminal nodes with mean around $1.25$ and standard deviation $0.55$. Also this section is divided in a data and likelihood, and a results subsections.

\subsubsection{Data and Likelihood}

Here, we apply the BART model with $m = 10, 20$ in the case of binary observations. The likelihood is the same reported by Equation \ref{eq:binary_lik} but it changes the expression of $p(\bx) = \Pr(Y = 1|\bx)$. In this case, this is given by

\begin{equation}
p(\bx) = \Phi(f(\bx)), \quad\text{where } f(\bx) = \sum_{j = 1}^{m} g(\bx, T_j, M_j),
\end{equation}
and $\Phi(\cdot)$ is the standard normal cumulative distribution function.The MCMC algorithm presented in Section 4 for gaussian models is modified for the binary case accordingly to \cite{chipman2010bart}.

Regarding the dataset, it originally comprises $101,766$ observations and $47$ variables (or covariates). The variables include physiological characteristics of the patient (e.g. gender, race, age, weight) as well as information on the history of the patient, such as number of visits or emergencies in the year prior to the encounter, and information relative to the encounter itself. For a detailed list of the features contained in the dataset and their description we refer to the website \footnote{\url{http://archive.ics.uci.edu/dataset/296/diabetes+130-us+hospitals+for+years+1999-2008}}. We decided to exclude the categorical variables that have a number of categories greater or equal to $10$, and the ones for which more than the $95\%$ of the observations have the same value; we kept all the numerical covariates. After this selection, we created dummy variables for the remaining categorical variables, which lefts us with $32$ features.

We divided the males ($45,918$) and females ($55,848$) and, for each group and prior under study, we fit BART models considering $m = 10$, and $20$ trees. For both male and females, and under both priors, we reached similar levels of accuracy ($\approx 62\%$) and increasing the number of trees did not provide any relevant advantage in terms of this. Therefore, given that the aim of this Section is to show that under the LB prior the MCMC algorithm explores shorter trees without loosing in terms of accuracy or likelihood, we only show the results for the males. The same results hold also for the female group. 

\subsubsection{Results}

For each prior, and for each number of trees $m = 10, 20$, we considered 10 independent chains of 1500 iterations with burn-in 500. In Appendix I we show the trace plots of the log-likelihood, missing rate and average number of terminal node per iteration for the different chains. We can see that even though the log-likelihood is still slowly increasing, this does not translate in a substantial improvement in terms of missing rate, which stays almost constant around $0.38$. Also, increasing the number of trees from 10 to 20, did not provide any gain in this regard, and therefore we can consider the algorithm to have converged. Comparing the results for the LB and CL priors, we see that the latter provides slightly higher log likelihood, but higher average number of terminal nodes. The gain in terms of log-likelihood does not provide a gain in terms of missing rate. 

\begin{figure}
\centering
\includegraphics[width=\linewidth]{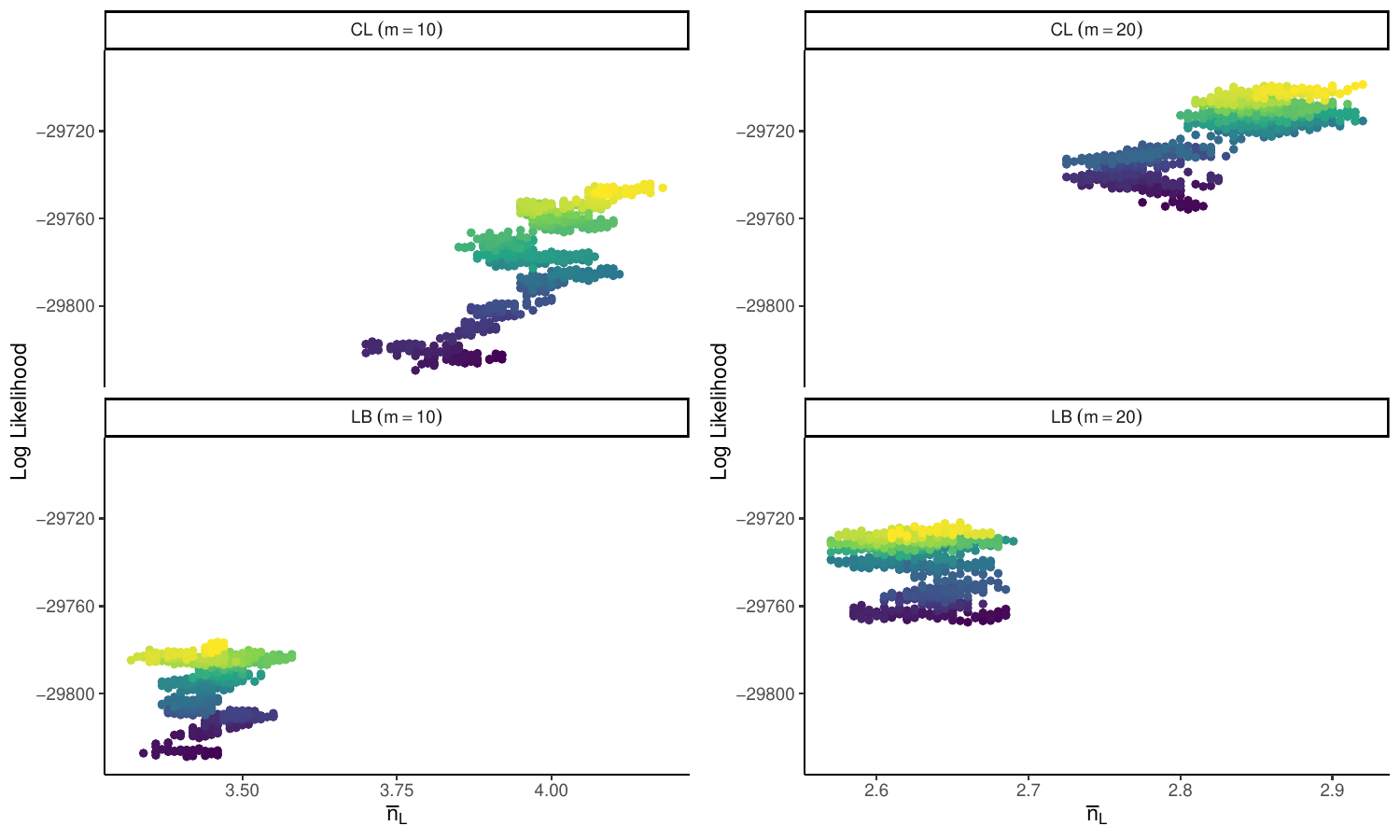}
\caption{Scatter plots of the average number of terminal nodes per iteration per chain versus the average log-likelihood per chain. The color indicates the iteration going from darker to lighter tones. Two BART models with different number of trees ($m = 10,20$) are considered, as well as, two priors for the tree topology (LB: loss-based with $\omega = 1.52, \gamma = 0.62$; CL: classic with $\alpha = 0.25, \beta = 2$). Different panels represents different combinations of prior and number of trees.}
\label{fig:13_diab_loglik}
\end{figure}

\begin{figure}
\centering
\includegraphics[width=\linewidth]{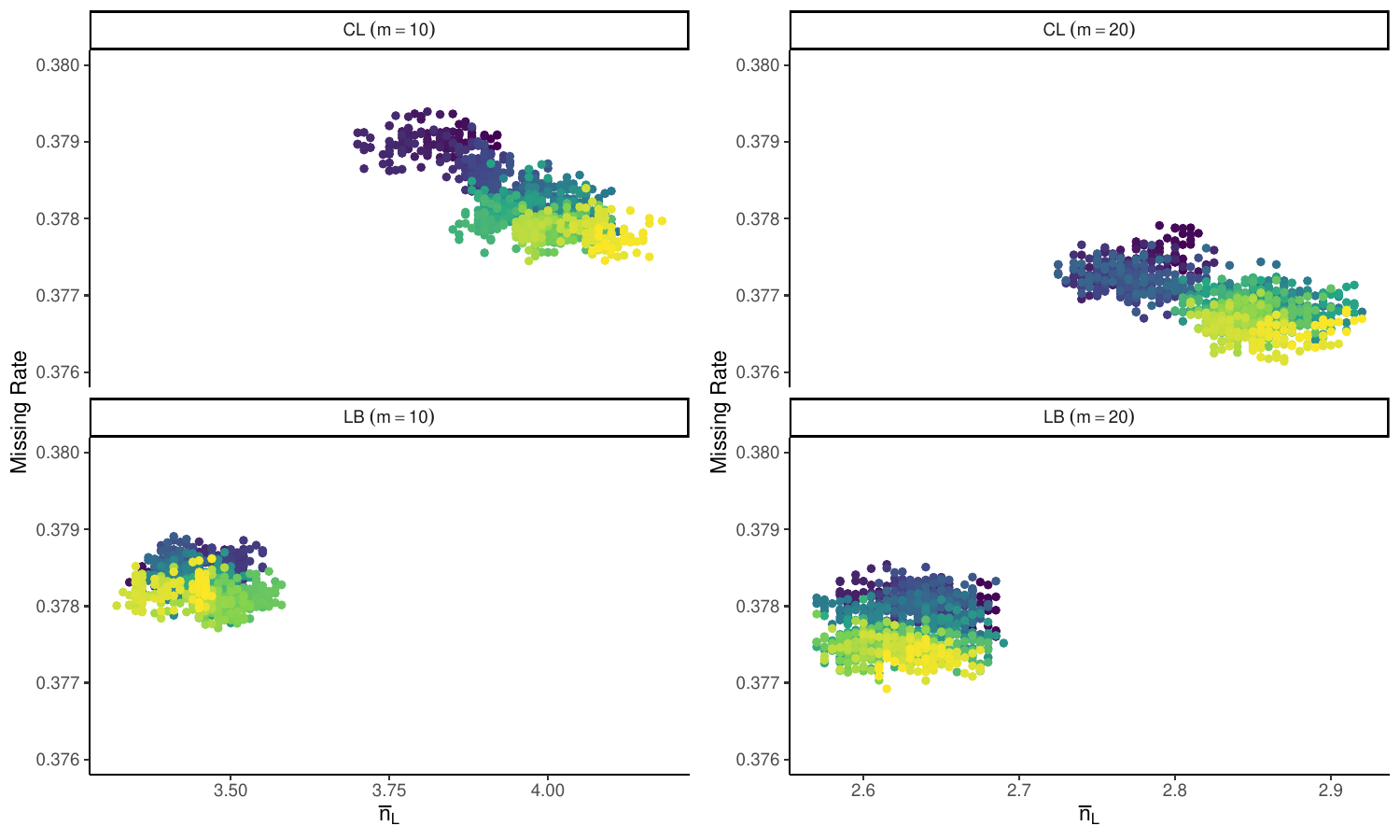}
\caption{Scatter plots of the average number of terminal nodes per iteration per chain versus the average missing rate per chain. The color indicates the iteration going from darker to lighter tones. Two BART models with different number of trees ($m = 10,20$) are considered, as well as, two priors for the tree topology (LB: loss-based with $\omega = 1.52, \gamma = 0.62$; CL: classic with $\alpha = 0.25, \beta = 2$). Different panels represents different combinations of prior and number of trees.}
\label{fig:14_diab_miss}
\end{figure}

The results for the different chains are summarised in Figures \ref{fig:13_diab_loglik} and \ref{fig:14_diab_miss}, showing the relationship between, respectively, average log-likelihood and missing rate per iteration per chain, with the complexity of the explored trees. The latter is quantified by the average number of terminal nodes per iteration per chain $\bar{n}_L$, which is obtained taking the average over the different chains of the average number of terminal nodes per iteration. The Figures show that, considering the LB prior, the MCMC algorithm explores shorter trees. Indeed, $\bar{n}_L$ is always below 3.6 (2.7) under the LB prior considering $m = 10$ (20) trees, while it is always over this value under the CL prior. This explains the small gain in terms of log-likelihood, which however does not translate in terms of missing rate which is basically the same under the two priors.

These results are strengthened if we look at the evolution of the log-likelihood or misspecification rate with the iteration indicated by the color of the points in Figures \ref{fig:13_diab_loglik} and \ref{fig:14_diab_miss}. It is clear that under the CL prior, as the number of iteration grows (more yellow), so it does the average number of terminal nodes, and therefore also the log-likelihood; indeed the points form a \emph{diagonal} cloud. On the contrary, under the LB prior, a growth in the number of iterations does not correspond to a systematic increase in tree complexity, despite an increase both in terms of log-likelihood and missing rate. This shows again that under the LB prior, the MCMC algorithm is more capable of optimising shorter trees to reach higher log-likelihood regions than under the CL prior that does that by also increasing the complexity of the trees.

\section{Discussion and conclusions}
\label{sec:conclusions}

In this article, we have applied the loss-based prior approach proposed by \cite{villa2015objective} to design a prior (LB) for the tree topology of BART and CART models. The obtained prior explicitly penalises for the number of terminal nodes ($n_L$) and the difference between left and right terminal nodes ($\Delta$) giving more control over the shape of the trees than the classical tree prior (CL) proposed by \cite{chipman1998bayesian} which also penalises for the skewness of the tree but indirectly. 

The prior we propose have a simple functional form and the parameters have a clear interpretation. Indeed, the prior on the number of terminal nodes is a Geometric distribution, while the conditional distribution of $\Delta|n_L$ is a discrete distribution with finite domain and known probability mass function. This facilitates setting the prior parameters in case of prior information on the tree structure, alternatively, we showed that default parameters settings can be found optimising the expected loss. Using synthetic and real data experiments we showed that the MCMC routine explores shorter trees under the LB prior than under the CL prior without loosing in terms of in-sample and out-of-sample prediction capabilities. We also explored the effect this has on the predictors that are used, and that exploring shorter trees provides greater separation between relevant and irrelevant predictors, providing better variable selection. This property makes the LB prior a good candidate to be used in combination with priors on the splitting rules designed to provide better variable selection such as the ones proposed by \cite{rockova2017posterior} and \cite{linero2018bayesian}. 

We have studied the differences in the results under the LB and CL prior considering applications of the CART and BART models on synthetic and real data. All the cases we have considered consistently show that under the LB prior shorter trees are visited, and despite this may lead to lower log-likelihood, this do not translate in better performance in in-sample and out-of-sample forecasting experiments. More specifically, we showed that considering LB and CL priors with similar prior mean and standard deviation for the number of terminal nodes, the LB prior penalises more heavily for the complexity of the trees and induces the MCMC algorithm to optimise more effectively shorter trees. This happened for CART models, on synthetic data simulated using a simple tree, and real data with limited number of observations (683) and predictors (9), as well as for BART models with a larger number of observations ($45,918$) and predictors (32). In all cases, we observe that, under the LB prior, the MCMC algorithm explores shorter trees than under the CL prior but with similar values of log-likelihood and missing rate. This is indicative of the fact that the LB prior penalises more for complexity (but not too much), which in turn induces the MCMC algorithm to find \emph{better} (high log-likelihood, low complexity) trees than under the CL prior. This results are encouraging because it may well have been that applying a larger penalty for complexity prevented the MCMC routine to visit trees in high log-likelihood region, which would have led to underperforming models. However, we observe that the higher penalty is well balanced, and the opposite happens. The diabetes data application also shows that this result holds also when increasing the number of trees in the BART model.

The loss-based prior approach described and used in this paper is not limited to the prior we have shown. Indeed, it could be used both to design different priors for the tree topology, or to design priors for other model components (e.g. the number of trees). More specifically, the LB approach depends on which tree statistics are used in defining the loss in complexity ($n_L$ and $\Delta$ in this article). For example, in early experiments we used the depth of the tree and a conditional distribution on the number of terminal nodes given the depth. This prior did not perform well in synthetic experiments and we had to change statistics. Despite the prior was not optimal, this testifies to the flexibility of the approach and the ability to target specific tree's statistics and potentially tailor the LB prior on the problem at hand. In the same way, this approach can also be used to design priors for other quantities. For example, it has already been used to perform variable selection \citep{villa2020variable} and therefore it could be used to design a prior on the splitting rules that penalises for the number of predictors used in a tree. Along the same lines, it can be used to design a prior distribution on the number of trees in BART. This would be highly beneficial because for now (w.r.t the authors knowledge) we do not make inference on this quantity which is considered as a tuning parameter. Having a prior on the number of trees would allow us to make inference on this quantity based on observed data rather than trying different values and subjectively picking the best one. In conclusion, we believe our approach can provide a valuable and fruitful addition to the CART/BART model literature with the potential of opening up new lines of research in this field.      

\section{Data availability statement}

The breast cancer data used in Section \ref{sec:real_data} has been 
collected by William HI. Wolberg, University of Wisconsin Hospitals,
Madison \citep{wolberg1990multisurface}, and already analysed by various authors. The dataset can be downloaded from the University of California Irvine repository of machine-learning databases \footnote{\url{https://archive.ics.uci.edu/dataset/15/breast+cancer+wisconsin+original}}. The code used to produce the analyses in Section \ref{sec:sim_study} and \ref{sec:real_data} as well as all the figures of the article is publicly available on GitHub \footnote{\url{https://github.com/Serra314/Loss_based_for_BART}}.

\section{Acknowledgments}
The authors are grateful to the Leverhulme Trust for funding this research (Grant RPG-2022-026).









\bibliographystyle{abbrvnat}

\end{document}